\documentclass[twocolumn]{aastex61}
\usepackage{color}
\usepackage{booktabs}

\begin{document}

\title{MODELED TEMPERATURE-DEPENDENT CLOUDS WITH RADIATIVE FEEDBACK IN HOT JUPITER ATMOSPHERES}
\author{Michael Roman}
\affiliation{Department of Phyiscs and Astronomy, University of Leicester, Leicester LE1 7RH, UK}
\affiliation{Department of Astronomy, University of Michigan, Ann Arbor, MI 48013, USA}
\author{Emily Rauscher}
\affiliation{Department of Astronomy, University of Michigan, Ann Arbor, MI 48013, USA}

\begin{abstract}
Using a general circulation model with newly implemented cloud modeling, we investigate how radiative feedback can self-consistently shape condensate cloud distributions, temperatures, and fluxes in a hot Jupiter atmosphere.  We apply a physically motivated but simple parameterization of condensate clouds in which the temperature determines the cloud distribution, and we evaluate how different assumptions of vertical mixing and aerosol opacity affect predictions.  We compare results from cases in which the aerosols are simply included in the last step of the simulation (i.e. post-processed) to cases in which clouds and their radiative feedback are actively included throughout the duration of the simulation. When clouds and radiative feedback were actively included, cloud cover decreased at equatorial regions and increased towards the poles relative to the post-processed solutions. The resulting phase curves also differed between the two approaches; the post-processed cloud simulations predicted weaker day-night contrasts in emission and greater eastward shifts in the maximum emission compared to the active cloud modeling. This illustrates the importance of cloud radiative feedback and shows that post-processing can provide inaccurate solutions when clouds are thick enough to provide significant scattering.
  
\end{abstract}

\section{INTRODUCTION}

Aerosols are likely common features of planetary atmospheres \citep{Lodders2010, Marley2013exoclouds} and may be seen as fundamental attributes of any significant, chemically-rich, atmosphere over an appropriate range of temperatures. Whether in the form of photochemical hazes or vapor condensate clouds, aerosols of various compositions are clearly prevalent in the range of environments within our own solar system \citep{[e.g] West1986, Gao2017clouds}. Given the multitude of possible atmospheric gases and the range of temperatures expected in other distant atmospheres, we may reasonably likewise expect a prevalence of clouds and hazes in most exoplanetary atmospheres.  

Considering their likely presence and largely undetermined characteristics, aerosols have been an attractive explanation for several anomalous characteristics in observations.  As a source of opacity, aerosols have been used to explain unexpectedly subdued spectral features in transmission spectra \citep{Gibson2012,Gibson2013,SingNature2016, Kreidberg2018} and weaker than expected thermal emission on the nightsides of hot Jupiter atmospheres \citep{Stevenson2014,Kataria2015,Stevenson2017,Mendoca2018}.  As a potential source of reflectance, inhomogeneous clouds have been proposed to explain the high albedo and possible asymmetries in the reflected light phase curves of several hot Jupiters \citep{Esteves2015phasecurves}, particularly the hot Jupiter Kepler 7b \citep{Demory2013,Hu2015, Webber2015, MnI2015, Parmentier2016}. 

Through scattering and absorption, clouds and hazes have the potential to alter the atmospheric energy balance, heating rates, and consequent temperature structure and winds \citep{Moreno1997heating, HengDemory2013}; this in turn can alter the environment in which aerosols form and modify observable quantities such as the flux and distribution of reflected and emitted radiation.  Evaluating the potential effects of clouds and hazes on these observables is thus important for adequately interpreting observations and characterizing exoplanetary atmospheres.

However, the complexity of the physical processes that govern clouds and hazes makes rigorous modeling at the resolution of a three-dimensional general circulation model (GCM) challenging.  The range of scales and processes involved requires significant parameterization in even the most advanced numerical weather prediction models \citep{bauer2015numericalmodels}, and tradeoffs must be made between accurately capturing the relevant physics and exploring a wide, largely uncharted parameter space.  This has led diversity of approaches to modeling aerosols in GCMs.  The most complex and computationally demanding atmospheric modeling of exoplanets has included self-consistent, spontaneous aerosol formation with modeled microphysics, feedback, and evolution of clouds in their predictions of winds, temperatures, and cloud distributions \citep{Lee_kitchensink2016, Lee_dynamical_clouds2017, Lines2018}.  Alternatively, simpler but still valuable approaches have instead computed temperatures for a clear atmospheres, and then used these results to predict clouds distributions by comparing temperatures to condensations curves \citep{Kataria2016,Parmentier2016,lewis2017cloudevo}, or additionally comparing fall rates to vertical winds \citep{Oreshenko2015}. This post-processing approach provides an expedient way for predicting distributions of clouds and evaluating their potentially observable consequences, but it necessarily neglects the effects of aerosols on the atmosphere.  

As mentioned, the presence of aerosols can alter the environment in which they form.  Since condensation and most chemical reaction rates and are dependent on the temperature, the radiative response of aerosols can shape the ensuing distribution of aerosols. This feedback may be of secondary importance if the clouds are thin (as \cite{Parmentier2016} concluded in their study), but they may be more consequential if clouds are thick.  \cite{Roman&Rauscher2017} modeled thick clouds of fixed distributions in the atmospheres of Kepler 7b, based on modeling of observed reflectivity by \cite{MnI2015}. \cite{MnI2015} had concluded that very high, thick clouds positioned along the western terminator were needed to reproduced the observed reflected light phase curves.  By including the double-gray radiative effects of these static, prescribed clouds in a GCM, \cite{Roman&Rauscher2017} showed that if clouds were assumed to be very thick, the radiative effects of clouds could significantly alter the heating rates, resulting in temperature fields that were inconsistent with the prescribed distribution of condensate clouds. As a consequence of these forced, self-inconsistent distributions, the insulated equatorial regions cooled less efficiently and emitted more flux on portions of the nightside than expected. Though extreme, that modeling illustrated the significance of ignoring the natural effects of radiative feedback on the cloud cover.

In the present study, we further investigate how radiative feedback can shape the condensate cloud distribution, temperatures, and observable fluxes in a hot Jupiter atmosphere.  We build upon the previous modeling of \cite{Roman&Rauscher2017} to include physically motivated parameterizations of condensate clouds that includes temperature dependence, akin to typical post-processing methods, but with radiative feedback effects throughout the duration of the simulation.  We discuss our methods for parameterizing clouds within GCM simulations in Section \ref{methods}. In Section \ref{Results}, we compare the results of simulations completed using post-processing techniques to those that included active radiative feedback for three different assumptions regarding cloud parameters. We show that the different approaches can result in significant differences in cloud cover and resulting reflected and thermal phase curves, as summarized in our conclusions in Section \ref{conclusions}.
 
\section{METHODS} \label{methods}

\subsection{The General Circulation Model}

To simulate temperatures, winds, and radiative fluxes within both clear and cloudy atmospheres, we used the GCM previously described in \cite{Roman&Rauscher2017} but significantly developed in its modeling of clouds.  Originally based on the Intermediate General Circulation Model of the University of Reading \citep{Hoskins1975}, the code was previously modified to model hot Jupiters \citep{MenouRauscher2009, RauscherMenou2010,RauscherMenou2012}. The GCM solves the primitive equations of meteorology using the spectral method of discretization in the horizontal and finite differencing in the vertical.  As described in detail in \citep{Roman&Rauscher2017}, the radiative transfer scheme applies a double-gray, two-streamed approximation of radiative transport that includes effects of aerosol scattering following \cite{Toon1989}.

For our simulations we chose to use the planetary parameters of Kepler-7b \citep{Latham2010}, a hot Jupiter for which specific inhomogeneous aerosol distributions have been proposed to explain the observed reflectances \citep{Demory2013, MnI2015}. The expected atmospheric temperatures of Kepler-7b span a range that includes the condensation curves of several abundant silicates, such that condensate clouds may form in the coolest regions while the minerals remain in the vapor state in the hottest regions (as discussed in Section \ref{clouddisc}); this provided a environment for potentially rich, heterogeneous cloud distributions. The choice of planetary parameters also allowed us to directly compare our results to those of our previous study, in which we modeled prescribed, fixed clouds on Kepler-7b \citep{Roman&Rauscher2017}.  As in our previous work, we neglect the expected magnetic effects that may potentially influence the circulation at these hot temperatures, in order to isolate role of radiative feedback and responsive temperature-dependent clouds in the simulation.  The model parameters used are listed in Table \ref{table:modparams}.

Initial conditions at each location assumed still winds and a temperature profile appropriate for the cloud-free, dayside atmosphere.  This initial temperature profile was computed using the analytical approximation of \cite{guillot2010radiative}, with absorption parameters chosen to best match the temperature profile modeled by \cite{Demory2013} using methods of \cite{Fortney2008unified}.

\begin{figure*}[th!]
\includegraphics[clip, width=\textwidth]{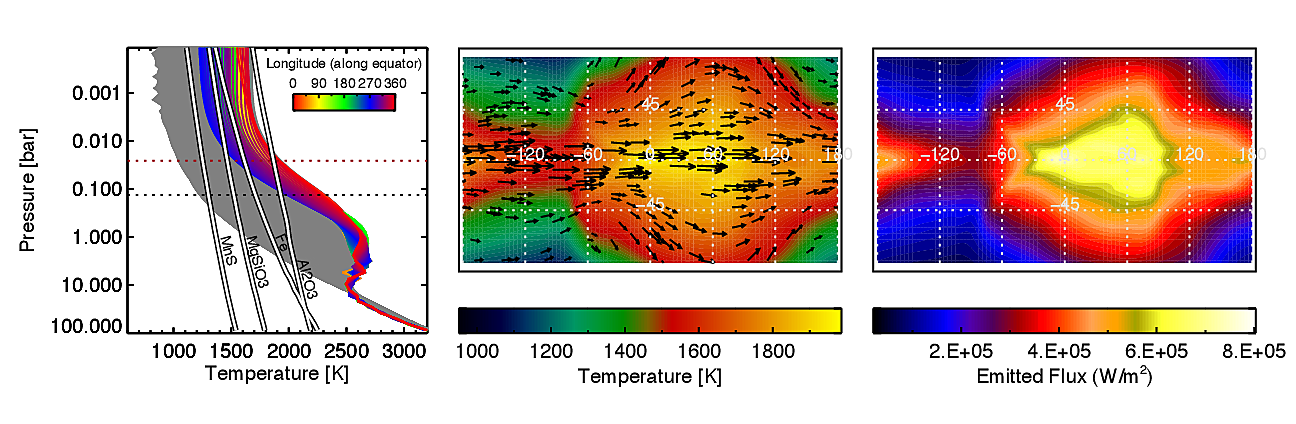}
\caption{Temperatures, winds, and emission from our simulation of a clear hot Jupiter atmosphere. (Left Panel) Temperature profiles for the cloudless atmosphere compared with condensation curves of selected condensibles. Equatorial locations at different longitudes are shown in color as indicated in the legend, while gray lines are from all other locations. For all plots, the substellar point defines zero latitude and longitude. The horizontal red and black dotted lines represent the pressure of the IR photosphere and where the two-way optical depth of visible light equals one, respectively, for a clear atmosphere.  Condensation curves are shown for MnS (alabandite), MgSiO${_3}$ (enstatite), Fe (iron), and Al${_2}$O${_3}$ (corundum), as included in the cloudy simulations, taken from Figures 1 and 2 of \cite{Mbarek&Kempton2016}. Clouds are expected to form where temperature profiles cross condensation curves. (Middle Panel) Mapped temperature field from our clear atmosphere at the 25 mbar pressure level, roughly corresponding to height of the infrared photosphere, along with wind vectors showing the direction and relative strength of winds at that height. Temperatures fields like this may be post-processed to predict cloud coverage. (Right Panel) Global map of the emitted thermal flux (integrated over all wavelengths) at the top of the model. In the absence of attenuating clouds, the emission map is determined by the temperature field alone.}
\label{fig:clear}
\end{figure*}

Results from a simulation of the clear model are shown in Figure 1. Our clear model exhibits several of the characteristics typical of many hot Jupiter simulations.  The synchronously locked dayside is hotter than the nightside, and there is strong equatorial jet that advects the hot spot to the east of the substellar longitude \citep{showman2009atmospheric,Dobbs-Dixon2010,RauscherMenou2010,Heng2011,Mayne2014, Chogroup2014}. Clear of any scatterers, the thermal emission from the planet mirrors this temperature pattern at the infrared photosphere.  

\begin{deluxetable*}{lccc}[th!]
 \centering
\tabletypesize{\footnotesize}
\tablecolumns{4} 
\tablewidth{4in}
 \tablecaption{ General Model Parameters}
 \label{table:modparams}
 \tablehead{
 \colhead{Parameter} & \colhead{Value} & \colhead{Units} & \colhead{Comment} }
 \startdata 
\it{    Orbital/Dynamical}\\
Radius of the planet, $R_p$ & $1.128\times 10^8$ & m & ref: Demory et al. 2011 \\
Gravitational acceleration, $g$ & 4.17 & m s$^{-2}$ & ref: Demory et al. 2011\\
Rotation rate, $\Omega$ & $1.49 \times 10^{-5}$ & s$^{-1}$ & assumed tidally synchronized\\\\
\it{    Clear Atmosphere Radiative Transfer}\\
Specific gas constant, $\mathcal{R}$ & 3523 & J kg$^{-1}$ K$^{-1}$ & assumed $H_2$ rich\\
Ratio of gas constant to heat capacity, $\mathcal{R}/c_P$ & 0.286 & -- & assumed Diatomic \\
Incident flux at substellar point, $F_{\downarrow \mathrm{vis}, \mathrm{irr}}$ & $1.589 \times 10^6$  & W m$^{-2}$ & ref: Demory et al. 2011\\
Internal heat flux, $F_{\uparrow \mathrm{IR}, \mathrm{int}}$ & 2325 & W m$^{-2}$  & from modeled T-profile\\
Visible absorption coefficient, $\kappa_{\mathrm{vis}}$  & $1.57 \times 10^{-3}$  & cm$^2$ g$^{-1}$ & constant, from modeled T-profile \\
Infrared absorption coefficient, $\kappa_{\mathrm{IR}}$  & $1.08 \times 10^{-2}$  & cm$^2$ g$^{-1}$  & constant, from modeled T-profile\\
\it{     }Pressure of $\tau_{vis}$ = 1 for two-way vertical path & $133$ & mbar & gas only, calculated from $\kappa_{\mathrm{vis}}$ and $g$\\
\it{     }Pressure of $\tau_{vis}$ = 2/3 for one-way vertical path & $177 $ & mbar & gas only, calculated from $\kappa_{\mathrm{vis}}$ and $g$ \\
\it{     }Pressure of $\tau_{IR}$ = 2/3 for one-way vertical path & $26 $ & mbar & gas only, calculated from $\kappa_{\mathrm{IR}}$ and $g$ \\\\
\it{    Model Resolution}\\
Vertical layers & 50 & --\\
Bottom of modeling domain pressure &  $\sim$100 & Bar \\
Top of modeling domain pressure & $5.7 \times 10^{-5}$ & Bar\\
Horizontal Resolution & T31 & -- & corresponds to  $\sim$48 lat $\times$  $\sim$96 lon\\
Dynamical Temporal Resolution & 4800 & time steps/day & \\
Radiative Transfer Temporal Resolution  & 600 & time steps/day & heating rates updated every 8 timesteps\\
Simulated Time & 2000 & planet days & \\
 \enddata
 \vspace{-0.8cm}
 \label{table:modparams}
\end{deluxetable*}

\subsection{Cloud Modeling within the GCM}
Whereas \cite{Roman&Rauscher2017} prescribed fixed aerosol distributions motivated by observations, for the present study we modified the code to allow for more realistic, mutable clouds with physically motivated distributions. The location of clouds depended on the modeled temperature field relative to condensation curves for different potential condensates. If a given temperature profile crossed a condensation curve, a cloud base was assumed to form at that location, and the layer was assigned scattering properties appropriate for the particular cloud species.  The visible opacity, infrared opacity, and vertical thickness of exoplanetary clouds are not well constrained by observations and theoretically dependent on assumptions regarding the abundances, particle sizes, condensation efficiency, scattering properties, and strength of vertical mixing, as discussed below. Given the large number of variables, we chose to limit our focus to a few cases that express different assumptions regarding overall cloud opacities and vertical mixing, with the goal of evaluating the significance of these uncertainties on the modeled atmosphere.  Then for each of these cases, we ran and compared simulations applying two different approaches to processing aerosols: in one approach, clouds were permitted to form, alter the heating rates, and respond throughout the duration of the run\textemdash{which we refer to as active cloud modeling}\textemdash and in the other, the cloud coverage was simply determined from the last iteration of a clear model\textemdash{which we refer to as post-processing, as detailed below.}

\subsubsection{Cloud Composition and Condensation Curves} \label{clouddisc}
We compared temperature profiles to condensation curves based on values in Figures 1 and 2 of \cite{Mbarek&Kempton2016}, in part reproduced in Figure \ref{fig:clear}. Though several species can condense at the relevant temperatures and pressures, only a few are expected to have abundances great enough to yield clouds of significant opacity; these include Al${_2}$O${_3}$ (corundum) and related aluminum bearing condensates, CaTiO${_3}$ (perovskite) and related titanium bearing condensates, Fe (iron), Mg${_2}$SiO${_4}$ (forsterite), MgSiO${_3}$ (enstatite), Cr (chromium), and MnS (alabandite), listed in order of decreasing condensation temperatures \citep{Parmentier2016, Wakeford2017}. Of these, for a solar composition atmosphere, MgSiO${_3}$ would produce the most condensate while CaTiO${_3}$ and Cr would produce the least \citep{Wakeford2017}. Given this, for simplicity and clarity we chose to include only MnS, Al${_2}$O${_3}$, Fe, and MgSiO${_3}$, which provide four distinct condensation curves that span the range expected atmospheric temperatures. We neglect all remaining possible condensates, even though small abundances could potentially contribute significantly to the scattering in the long path-lengths nearing the limb.  We note that despite its relatively meager mass, CaTiO${_3}$ was determined by \cite{Parmentier2016} to be one of the more important species due to its high condensation temperature and high albedo.  Al${_2}$O${_3}$ has a similarly high condensation temperature and much greater mass ($\sim$ 13 $\times$) compared to CaTiO${_3}$, but \cite{Parmentier2016} found that Al${_2}$O${_3}$ was a less effective scatterer when particles sizes were greater than 0.1 $\mu$m.  In our case, we investigate only slightly larger particles (radius of 0.2 $\mu$m) and include multiple cloud species simultaneously in all our simulations, with scattering parameters weighted by the opacity, meaning that the contribution of CaTiO${_3}$ would be diminished by the much more massive Al${_2}$O${_3}$. The reality is likely more complex, as the contribution of each species critically depends on how different species combine to form aerosols within a cloud, which is highly uncertain.  Nonetheless, the potential role of additional species will be explored in future modeling.

\subsubsection{Cloud Opacity and Scattering Parameters} \label{scatparamdisc}
The optical thickness of aerosols formed within a given layer were based on the following assumptions.  In general, the opacity was taken as a function of the gaseous abundance and cloud particle properties. We assumed the minor gases were well-mixed with uniform, solar abundant mole fractions \citep{Burrows&Sharp1999}.  The potential mass of aerosols in each of the 50 vertical layers (logarithmically spaced in pressure), was assumed proportional to the mass of the condensing gas in that layer.  For each layer, we computed this component gas mass using the mole fraction and the molecular weight relative to the mean atmospheric weight (assumed Jovian), such that  

\begin{equation} \label{eq1}
{m_g}=\frac{\Delta P}{g} {\chi_g} \frac{\mu_g}{\overline{\mu}}
\end{equation}

\noindent where ${\Delta P}$ is the change in pressure across the layer, ${g}$ is the gravitational acceleration, ${\chi_g}$ is the mole fraction of the condensible gas species, ${\mu_g}$ is the molecular weight of the gas species, ${\overline{\mu}}$ is the mean molecular weight of the atmosphere, and ${m_g}$ is the resulting mass of the component gas species in the layer.

The mass of each component gas species was then converted to an aerosol optical depth via the expression 

\begin{equation} \label{eq2}
{{\tau_a}}=\frac{3 {m_g} {Q_e}{f}}{4 r {\rho}}
\end{equation}

\noindent where ${r}$ is the aerosol particle radius, ${Q_e}$ is particle scattering extinction efficiency, ${\rho}$ is the particle density, ${f}$ is the fraction of the particular gas that actually forms condensate, and ${\tau_a}$ is the resulting aerosol optical thickness of the layer. For a uniform mole fraction and assumed ${f}$ and scattering properties, combining Eqs.(\ref{eq1}) and (\ref{eq2}) yields a uniform aerosol optical depth per bar of pressure within the cloud. 

The fraction of each particular gas that becomes part of the condensate cloud would in general depend on the local temperature and equilibrium vapor pressure of the gas, as well as the efficiency of nucleation and condensation \citep{bohren2000atmospheric}; the amount that remains condensed within the cloud on some appreciably long time scale would in part depend on the vertical mixing and microphysical processes that determine rainout.  To get a sense of appropriate values for ${f}$, we first evaluated the partial pressure of MgSiO${_3}$ versus its equilibrium vapor pressure for temperature profiles taken from our clear GCM simulation. All the excess vapor pressure was converted into a mass and compared to the simple mole fractional mass given by Eq.(\ref{eq1}). To account for just the deviation from a proper saturation vapor pressure calculation, we found a value of ${f}$ ranging between 0.7 and 0.9 is appropriate.  Availability of potentially limited condensation nuclei and rainout would serve to reduce this fraction further by an unknown amount.  We accept that in reality, a host of physical processes beyond the scope of this model would complicate this picture, and combined with significant uncertainties in all the parameters in Eq. (\ref{eq2}), the total opacity of the cloud is widely uncertain. Accordingly, we viewed ${f}$ as an unconstrained parameter that was tuned to explore a range of values.

As a practical matter, we found values of ${f}$ $\gtrsim$0.33 caused computational difficulties, as the resulting high optical thicknesses (exceeding $10^5$ per bar) produced extreme heating rates and numerical instabilities.  We note that \cite{Lines2018} described a similar problem with high heating rates in their modeling, which they addressed by placing initial upper bounds on the heating rates while allowing the atmosphere to gradually adjust. A similar approach may be taken in our future work, but in the present case, given the wide range of uncertainty and the simplicity of our modeling, we simply limited our thickest cloud models to values of ${f}$=1/10.  We also wished to examine how significant radiative feedback was for relatively thin clouds, and so we ran simulations with  ${f}$=1/100. For the assumed particle properties, these corresponded to still hefty total optical thicknesses over 14,000 and 1,400 per bar, respectively, in our visible band, with infrared cloud optical thicknesses further dependent on relative extinction at longer wavelengths as discussed below.

Common scattering parameters for each cloud composition were computed from Mie theory at 650 nm and 5 microns, using the indices of refraction from \cite{KitzmannHeng2018}. The two wavelengths were chosen to represent the visible and infrared channels of the double-gray radiative transfer modeling used in our GCM, as discussed in Section \ref{2xgraydisc}. The parameters included the single scattering albedo $\varpi_{0}$, which defines the fraction of incident light scattered by each particle with values ranging between one (conservative scattering) and zero (fully absorbing); the asymmetry parameter \textit{$g_0$}, which is related to the scattering phase function and indicates whether particles tend to scattered more isotropically (values approaching zero) or preferentially in forward or backward directions (approaching 1 and -1, respectively); and the extinction efficiency \textit{$Q_e$}, which we use to relate the particle size and abundance to an optical depth. 

Since we did not model particle growth and evolution, we were forced to assume a mean particle size and distribution for our Mie calculations.  We chose a log normal size distribution with an effective mean particle radius of 0.2 micron and a variance of 0.1 micron based on inferences of small particles sizes in observations \citep{MnI2015,Kreidberg2018}, previous exoplanet modeling \citep{Parmentier2013,Lines2018}, and distributions in terrestrial clouds \citep{lopez_1977lognormal}. These small particles scatter more efficiently in the visible than the IR resulting in higher cloud opacities in the visible relative to the thermal.  We note in general that the particle sizes are likely functions of location and height in the atmosphere \citep{Parmentier2013, Lee_dynamical_clouds2017, Lines2018}, and the particle distribution may likely be more complex and even bimodal \citep{powell2018formation,Lines2018}, but the simple approach adopted here helps to cleanly isolate the consequence of the different vertical distribution and opacity assumptions.

The scattering properties applied in our simulations are listed in Appendix Table \ref{table:scatparams}. In locations where multiple clouds form at the same level, the scattering parameters ($\varpi_{0}$, $g_{0}$) are weighted based on the optical depth of each species.  The total aerosol optical depth was taken as the sum of component condensates, each of which has a distribution dependent on its respective condensation curve.

\subsubsection{Cloud Vertical Extent}
With the potential optical depth of any given layer set by the above assumptions, we looked to parameterizing the vertical extent of the cloud.  Though the base pressure of the cloud is reasonably determined by temperature profiles (i.e. where the condensation curves and temperature profiles cross), the vertical extent of the clouds above the base would at least depend on the relative strength of the vertical mixing versus sedimentation \citep{Ackerman&Marley1989, GaoMarley&Ackerman2018}. Vertical mixing rates are not well constrained, and modeling estimates of parametrized eddy diffusivity values vary by orders of magnitude \citep{Moses2013, Parmentier2013, Agundez2014}. If particle sizes are modest or vertical mixing is relatively weak, we would expect a compact condensate cloud layer, as is typically inferred to exist in giant planets of our solar system \citep[e.g. the ammonia clouds on Jupiter see][]{Rossow1978,Moses2013,Ackerman&Marley1989,Banfield1998a}; however, dynamical modeling suggests that vertical mixing in hot Jupiters is vigorous and particle sizes are small, permitting vapor and clouds to be lofted to sub-millibar pressures, if temperatures permit, somewhat irrespective of the pressure of the cloud's base \citep{Parmentier2013, Lines2018}. Whether clouds are compact or extended is significant because each case would yield different top-of-the-atmosphere albedos, depending on the condensation temperatures and intrinsic reflectance of each cloud species, and this would affect the predicted patterns of heating and reflectance.

To evaluate the consequences of the vertical extent, we chose to model both compact and extended cases. For the compact cloud cases, the constant tau per per cloud opacity is truncated after five vertical layers ($\sim$1.4 scale heights) beyond its base. In extended cloud cases, we allowed the cloud to extend up to 0.1 mbar, regardless of the base pressure. Either way, we maintained the previously discussed criteria, such that clouds only formed within these vertical domains if the temperatures permitted, and when formed, each species of cloud had a uniform aerosol opacity per bar determined by the relative abundances. We ignored the scenario in which volatiles may be limited in vertical extent by means of a cold-trap, as explored by \cite{Parmentier2013, Lines2018}, since our warmer temperature structure and choice of condensates precluded this scenario.

\subsection{Accuracy of the Double-Gray Approximation} \label{2xgraydisc}
Our GCM modeling uses a double-gray radiative transfer approximation, which effectively separates the spectrum and its processing into just two fundamentally different regimes \textemdash a visible regime characterized by scattering and absorption of stellar light entering the top of the atmosphere, and a thermal regime defined by the emission, scattering, and absorption of intrinsic thermal radiation originating from within the atmosphere.  This approximation has the strength of being computationally faster and simpler than more rigorously accurate schemes that evaluate multiple bands of different frequencies, while still capturing the essential physics necessary for computing approximate heating rates to first order. It has been used in previous hot Jupiters models \citep[e.g.][]{Dobbs-Dixon2010,Heng2011,RauscherMenou2012} where it has been shown to yield thermal structures and induced circulations similar to those produced using multi-spectral radiative transfer.

However, the accuracy of the double-gray approximation when applied to a cloudy atmosphere requires a closer look, as scattering within clouds is strongly wavelength-dependent. Of greatest concern is the variation in cloud opacity with wavelength, characterized by the scattering extinction efficiency. This parameter determines the effective mean cross-sectional area and consequent optical thickness of a cloud as a function of wavelength. In general, the extinction efficiency drops significantly with increasing wavelengths once the particle's physical size becomes small relative to the wavelength \citep{hansen_travis_1974,KitzmannHeng2018}. For expected particle sizes of 0.1 micron or more, variation across the visible spectrum is expected to be small and clouds may be approximated as gray scatterers; however, in the infrared, the extinction trend becomes significant.  Clouds that are optically thicker in the visible become optically thinner in the infrared and essentially transparent to radiation at longer wavelengths. In a double-gray approximation, this trend across the infrared is neglected, such that a single cloud opacity is applied from the near-IR to the far-IR. Depending on the values chosen, this leads to thermal scattering that is too slight at shorter wavelengths and too large at longer wavelengths. To some extent, this is mitigated by the fact that blackbody emission also diminishes significantly with wavelength, but the effect can be most significant near the blackbody peak.

With this in mind, we attempted to choose a representative wavelength by minimizing the error in fluxes produced by a gray cloud versus a cloud with proper wavelength dependent extinction. To do so, we evaluated the Planck function for relevant atmospheric temperatures and calculated the total wavelength-integrated transmission through an overlying attenuating layer with wavelength dependent extinction for 0.2 $\mu$m radii particles based on the functional fits to extinction efficiencies for MgSiO${_3}$ from \cite{KitzmannHeng2018}.  Then we calculated the transmission assuming only gray extinction, evaluated at a different wavelength in each case that covered a large grid of possible values.  Calculations were then repeated for different values of the layer optical depths and underlying blackbody temperatures spanning the range found in our models. We found that for each case, a single extinction efficiency could approximately reproduce the total integrated emission, but the precise value was highly dependent on the blackbody temperature and layer optical depth. It was also dependent on the extinction curve, which is itself highly dependent on the assumed particle size, particle size distribution, and indices of refraction.  We can only broadly conclude that a single extinction in the IR that is chosen to be roughly 10$^{-2}$ to 10$^{-3}$ of that in the visible provided the least errors in transmission between layers in these simple gray versus non-gray studies. In our model set-up, this corresponds to evaluating the extinction efficiencies at wavelengths of roughly 4-6 microns. Furthermore, any single value applied to all situations will yield incorrect fluxes that are generally within a factor of a few, but can be off by an order of magnitude in the coldest, cloudiest locations, compared to a non-grey calculations. How this translates to errors in the overall heating rates and emission in the context of a global circulation model is more complicated and requires a full side by side grey vs non-grey GCM treatment, which should be addressed in future work.

Based on the above tests, we chose to simply evaluate the infrared scattering parameters at a wavelength of 5$\mu$m and caution that the double-gray framework introduces artificial limitations that can lead to systematic errors in the cloudiest regions. In the visible, where the wavelength dependence is less significant, we chose to evaluate scattering at 650 nm, since this value falls roughly in the middle of the \textit{Kepler} bandpass.

\subsection{Active Clouds vs Post-Processed Clouds}
For each of the cloudy cases, we calculated the cloud distributions using two different approaches. We refer to our first approach as the \emph{active cloud} modeling approach, in which clouds were permitted to form and evolve throughout the entire duration of the simulation.  At each radiative time step, the computed temperatures were compared to the condensation curves; if temperatures fell below the condensation curve, the layer was assigned scattering properties appropriate for the particular cloud species.  Since clouds scattered and/or absorbed radiation, they had the potential to alter the heating rates and temperature fields; and since the cloud distributions were dependent on the temperature fields, the dependence served as a simplified feedback mechanism for clouds within the model.  

We then ran simulations using an second approach, which we refer to as \emph{post-processing}. In these cases, clouds were permitted only at the last time step of an otherwise clear simulation, though applying the very same criteria for determining aerosol distributions as in the active cloud approach. Instantaneous aerosol distributions and fluxes were then computed for a single time-step so as to prevent any alterations of the temperature fields due to radiative feedbacks. The aerosol distributions, temperatures, fluxes, and phase curves for these post-processed results were compared to the active cloud results.   

\subsection{Modeling Summary}

In summary, we focus on four representative models, exploring two different vertical extents\textemdash{compact and extended}\textemdash and two different cloud opacities\textemdash{relatively thin and thick}:

\begin{enumerate}
    \item Compact cloud of lesser opacity: the cloud is truncated after five vertical layers ($\sim$1.4 scale heights) beyond its base, with an optical thickness per bar set by assuming only 1$\%$ of the vapor condenses (i.e. f=1/100), corresponding to a maximum potential optical thickness per bar of 1,409 in the visible.
    \item Extended cloud of lesser opacity: potential opacity as above ($\tau$ = 1,409 per bar), but the cloud is allowed to extend up to the 0.1 mbar height, regardless of base pressure.
     \item Compact cloud of greater opacity:  the cloud is truncated after five vertical layers ($\sim$1.4 scale heights) beyond its base, but now 10$\%$ of the vapor condenses (i.e. f=1/10), corresponding to a maximum potential optical thickness per bar of 14,092 in the visible. 
      \item Extended cloud of greater opacity: potential opacity as above ($\tau$ = 14,092 per bar), but the cloud is allowed to extend up to the 0.1 mbar height, regardless of base pressure.

   \end{enumerate}           

In each of the above cases, we essentially defined the permitted vertical distribution and thickness of the cloud, but the temperature field determined whether a cloud was actually realized within these permitted domains.  Cloud species included MnS, Al${_2}$O${_3}$, Fe, and MgSiO${_3}$, and aerosols were assumed to have a log normal size distribution with the effective mean particle radii of 0.2 $\mu$m with a variance of 0.1 $\mu$m in all cases. For clarity, we will refer to the lesser opacity cases as \textit{thin}, and the greater opacity cases as \textit{thick}, but we emphasize that both cases have integrated optical thickness well above unity.

All simulations were run for 2000 planet days of model time, which we found was long enough to ensure that the presented temperatures, winds, fluxes, and cloud distributions displayed no significant changes with additional time.   As the temperatures in the model changed and converged towards quasi-steady state solutions, so did the aerosol distributions. The described scheme is a simplification, as it ignores the affect of local vertical velocities, chemical disequilibrium, inhomogeneities that would result from rain-out, latent heating, and many other physical processes, but it provided a simple and very efficient means of roughly mimicking plausible aerosol distributions at each time step within the GCM.  In turn, this allowed us to assess the importance of aerosol radiative feedback in our simple cloud modeling. By doing so, we could ascertain when simple post-processing is sufficient in characterizing the cloud distribution and observables, and when including radiative feedback can produce significant changes. 

\section{RESULTS} \label{Results}

\begin{figure*}[th!]
\includegraphics[clip,trim= 0in 6in 0.in 10.8in, width=\textwidth]{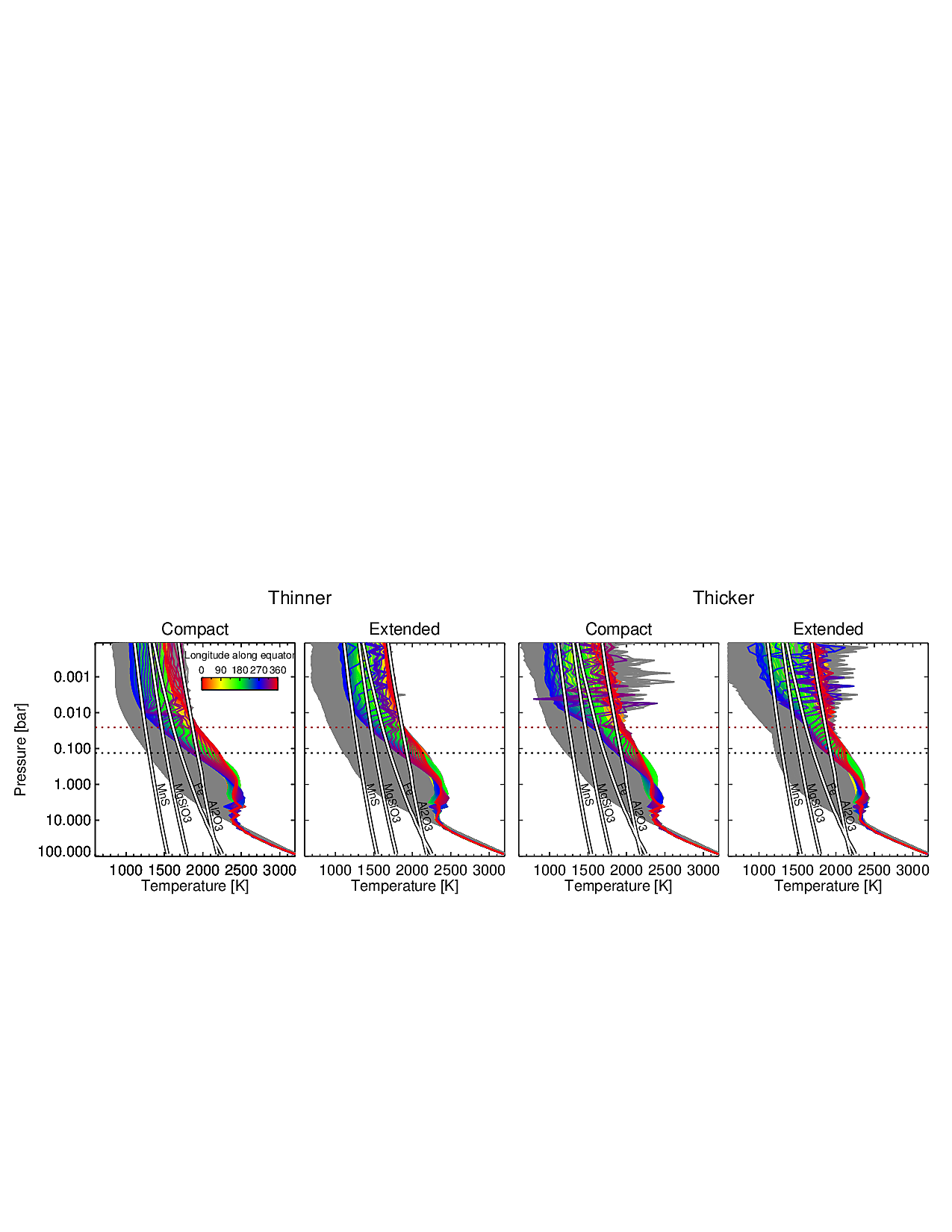}
\caption{Modeled self-consistent temperature profiles for four different cloudy cases along with condensation curves for MnS, MgSiO${_3}$, Fe, and Al${_2}$O${_3}$ from our GCM. Equatorial locations at different longitudes are shown in color as indicated in the legend, while gray lines are from all other locations. The horizontal red and black dotted lines represent the pressure of the thermal photosphere and where the two-way optical depth of visible light equals one, respectively, for a clear atmosphere. (Left Panel) Temperature profiles computed while including a compact clouds of relatively low potential cloud opacity ($\tau$=1,409 per bar), (Middle-Left Panel) clouds of low opacity but extended vertical distribution (up to 0.1 mbar), (Middle-Right) compact but optically thicker with $\tau$=14,092 per bar, (Right Panel) and clouds both thick ($\tau$=14,092 per bar) and extended, as discussed in the text.  Compared to the clear case (Figure 1), warmer temperatures and strong perturbations are evident across much of strongly irradiated dayside (red curves), where temperatures hover around the Al${_2}$O${_3}$ condensation curve.  At greater pressures, from several hundred millibars to 10 bars, the temperatures are cooler as less light penetrates through the clouds down to these depths. The thicker cloud models show more pronounced features including lower temperatures at the top of the model over the poles and nightside with warmer daysides.  While clouds do introduce abrupt changes in heating rates at cloud levels, the spikes in the profiles are exaggerated by the limited vertical resolution of the model and should be regarded as numerical artifacts.}
\label{fig:tprofs}
\end{figure*}

Using condensation curves and self-consistent modeled temperatures for a hot Jupiter, we computed idealized cloud distributions for a combination of different condensible species. To present our results, first we will examine where clouds form in context of the temperature fields and how their distributions differ depending on whether radiative feedback process are included or not. Then we look at the role of each cloud type, how each of these clouds contributes to observable reflectance on the dayside and reduced emission on the nightside.  Finally, we discuss how these differences, altogether, would affect the temperatures and consequent observable quantities, such as the emitted and reflected light as a function of orbital phase.

\subsection{Computed Temperatures}

In our simplified modeling scheme, the location of clouds were determined by the temperature field and assumptions regarding their vertical extent. For post-processed results, the temperature field was taken from the clear atmospheric result shown in Figure \ref{fig:clear}. For our active cloud results, temperatures were self-consistently computed with clouds present and allowed to adjust as clouds themselves adjusted throughout the duration of the simulation. 

\begin{figure*}[th!]
\includegraphics[clip, trim=.3in 4.8in .1in 3.6in,width=\textwidth]{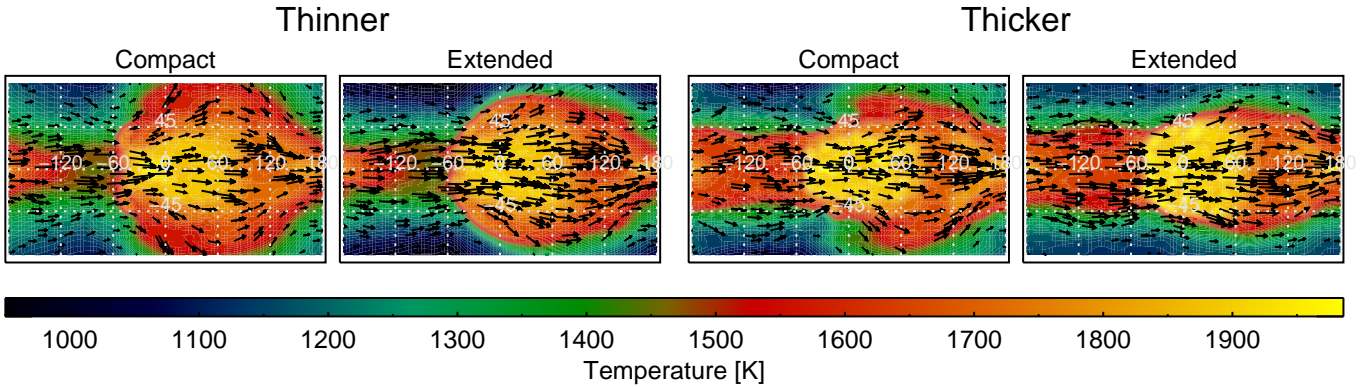}
\caption{Mapped temperature fields from our four active cloud models: the thin and compact cloud (left), thin and extended active cloud (middle-left), thick and compact cloud (middle-right), and thick and extended cloud (right) models, at the 25 mbar pressure level. The central latitude and longitude (i.e. 0$^{\circ}$, 0$^{\circ}$) marks the substellar point. The 25 mbar level roughly corresponds to the infrared photosphere for the clear atmosphere, where a majority of the emission to space originates. Compared to the clear case shown in Fig \ref{fig:clear} (shown with the same scaling and color bar), the high latitudes are cooled by clouds, particularly in the extended cases.  In the thicker cloud cases, the strong absorption and scattering heat the substellar regions to temperatures exceeding those in the clear atmosphere.}
\label{fig:tfields}
\end{figure*}

By means of scattering and absorbing radiation, the clouds' effect on the heating rates can be seen in the resulting temperature profiles and curves shown in Figures \ref{fig:tprofs} and \ref{fig:tfields}, in comparison to Figure \ref{fig:clear}. Though the magnitude of excursions in the jagged profiles are largely the numerical consequence of under resolving the sharp heating and cooling changes, resulting patchiness in the cloud cover unsurprisingly introduces physical variability in the modeled temperatures.

The warm equatorial jet and eastward shift of the hottest temperatures, as typically seen in models of hot Jupiters \citep{showman2009atmospheric,Dobbs-Dixon2010,RauscherMenou2010,Heng2011,Mayne2014}, are still found in cloudy simulations; however, clouds did significantly alter the heating rates and consequent temperature profiles when actively included. 

As the visible aerosol scattering dominated over the thermal scattering, the atmosphere was cooler at depth compared to the clear model (several hundred degrees at pressures of a few bars). This was especially true for the extended cloud models, where deep temperatures were colder than even the thicker-compact case. The cooling was simply due to the optically thick clouds' ability to block and reflect stellar radiation, preventing it from penetrating as deeply. More light is consequently scattered into space or absorbed above the cloud layers, leading to regions of both warmer and cooler temperatures that vary greatly by location. This effect was unsurprisingly greatest for the extended clouds because they occupy more layers and hence have a greater integrated optical thickness. At the poles, the colder temperatures allowed cloud bases to form deeper, and so there was a maximum difference between the realized vertical ranges of the compact and extended clouds.  In general, the extended clouds had a greater effect on the temperatures than the compact clouds.

High in the atmosphere, regions near the substellar point were warmer while regions along the western terminator, poles, and nightside were cooler compared to the clear case. This was a consequence of the chosen cloud properties and the radiative feedbacks they support.  MgSiO${_3}$ had a lower condensation temperature and formed thick, reflective clouds, and when these clouds formed along the cool western terminator or polar regions, they further cooled the atmosphere by scattering visible light (see Figs \ref{fig:compactcom} and \ref{fig:extendedcom}). 

In contrast, Al${_2}$O${_3}$, had higher condensation temperatures and formed more absorbing cloud particles; therefore, it was able to condense on the dayside, where particles absorbed more radiation and heated the atmosphere further. Heating from absorption continued until temperatures exceed the condensation temperature and the cloud vaporized, reducing the heating rate and causing the temperature to once again fall below the condensation curve.  This simple feedback caused the atmospheric profiles near the sub-stellar point to adjust to the Al${_2}$O${_3}$ condensation curve.  So, despite its relatively humble abundance, the Al${_2}$O${_3}$ had a pronounced effect on the dayside temperatures due to its relatively modest single scattering albedo and appropriately warm condensation temperature.  Other trace condensates could plausibly pay equivalent roles in other atmospheres, just as unaccounted condensates of high albedo could plausibly counter this effect (such as CaTiO${_3}$, which \cite{Parmentier2016} cites as a significant scatterer). In reality, the picture would likely be complicated by rainout, inhomogeneities, vertical motions, and latent heat, but such a simple feedback mechanism can potentially still play a part in shaping a range of dayside atmospheric temperature profiles. 

\subsection{Computed Cloud Distributions}

\begin{figure*}[bh!]
\includegraphics[clip, trim=0in 3in 0in 4in,width=\textwidth]{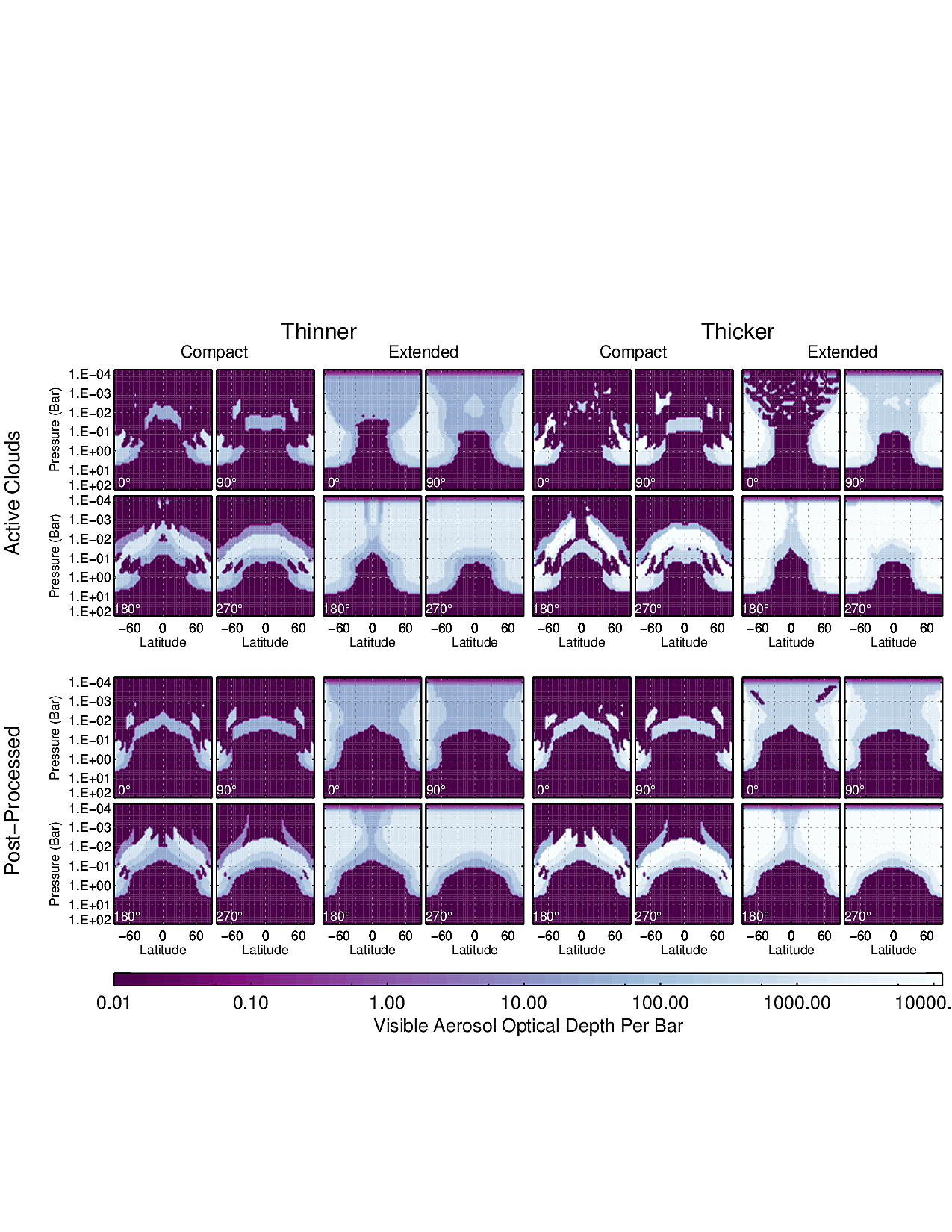}
\caption{Aerosol optical depth per bar for the compact thin cloud (left), extended thin cloud (middle-left), and compact thick cloud (middle-right), and extended thick cloud (right) models, computed using both active (top) and post-processed (bottom) approaches. Values are expressed at 650 nm, over our full domain of latitudes and pressures, at four different longitudes: 0$^{\circ}$, 90$^{\circ}$, 180$^{\circ}$, and 270$^{\circ}$, as indicated in the bottom left corner of each panel.  Radiative feedback causes greater clouds coverage at the poles and broken cloud coverage on the strongly irradiated dayside, particularly for the thicker cloud cases; the same effect is absent in the post-processed solutions.}
\label{fig:tauxsec}
\end{figure*}

\begin{figure*}[th!]
\includegraphics[clip, trim=0in 3.5in 0in 3.8in,width=\textwidth]{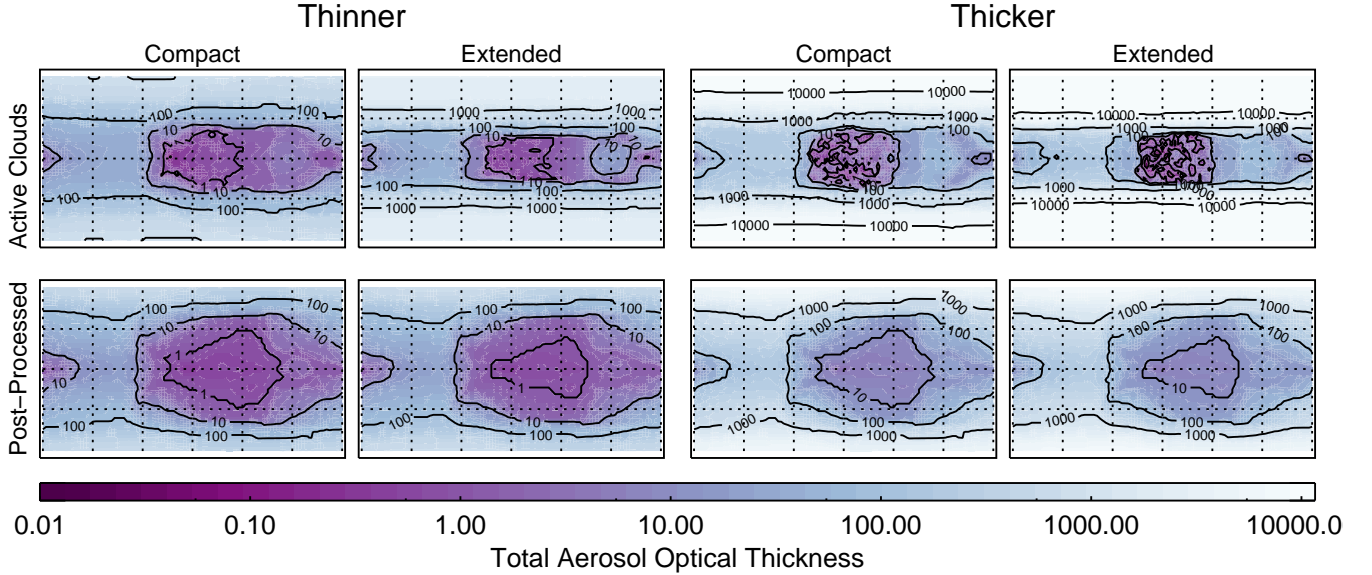}
\caption{The mapped integrated aerosol optical depth of the thin compact cloud (left), thin extended cloud (middle-left), and thick compact cloud (middle-right), and thick-extended cloud (right) models, resulting from active (top) and post-processed (bottom) approaches. The central latitude and longitude (0$^{\circ}$, 0$^{\circ}$) is the substellar point. The active clouds show greater optical depths at high and mid latitudes and broken cloud coverage near the substellar point compared to the post-processed cases. The inclusion of thermal radiative feedback caused the poles to become colder and cloudier due to self shadowing, while at the substellar point, feedback caused transient thin patches to develop in the strongly heated, absorbing clouds.}
\label{fig:taumap}
\end{figure*}

Figures \ref{fig:tauxsec} and \ref{fig:taumap} show the computed cloud distributions and how they differ between models given the different assumptions and implementations.  Figure \ref{fig:tauxsec} shows the vertical distribution of aerosols (optical thickness per bar) in our visible channel as a function of latitude for different values of the longitude.  Figure \ref{fig:taumap} shows the total vertically-integrated optical depth mapped in latitude and longitude  

Clouds formed deeper at high latitudes in all cases. Since most of the aerosol opacity occurs in the thicker atmosphere near the base of the cloud, the total integrated optical thickness was most sensitive to deeper clouds. For the same reason, when clouds were post-processed, the differences in the total integrated optical thickness between the compact and extended clouds were not large, since the warmer temperatures of the clear atmosphere precluded deeper clouds.  The deepest clouds formed only when the overlying clouds were actively able to cool the underlying atmosphere,  thus dramatically increasing the base pressure and total optical thicknesses. 

The difference between the active and post-processed modeling techniques was most significant on the dayside. The strong heating and cloud radiative feedback caused temperatures to fluctuate around the condensation temperature for Al${_2}$O${_3}$ near the substellar point, as seen in the temperature profiles.  This resulted in the patchy cloud cover on the dayside when active clouds of small particles were considered.  The effect is only on the dayside and is completely absent in the post-processed, clearly indicating that this was due to heating in the visible.  

These comparison illustrate how radiative feedback can alter the cloud distributions within our model, even when clouds are relatively thin.  The post-processed thin-cloud case matched the active cloud result best, but the results unsurprisingly diverged with increasing cloud opacity.

While we find that visible scattering dominates on the dayside, we note that this is sensitive to the assumed small particle size and choice of cloud compositions. Test simulations show that larger particle sizes and different scattering parameters can lead to more significant infrared scattering by clouds as their opacity is enhanced at thermal wavelengths. Thermal radiative feedback can affect the cloud cover by eroding the base of the cloud layer, effectively raising the cloud base and reducing the overall optical thickness.  As clouds are expected to have larger particles at the cloud base \citep{Parmentier2013, Lines2018}, this effect may be significant in shaping the cloud base and should be considered in future investigations.

\subsection{Component Cloud Contributions to Dayside Reflectances and Nightside Attenuation}

\begin{figure*}[th!]
\centering
\includegraphics[clip, trim=1.5in 5.5in 3.2in 2in, width=\textwidth]{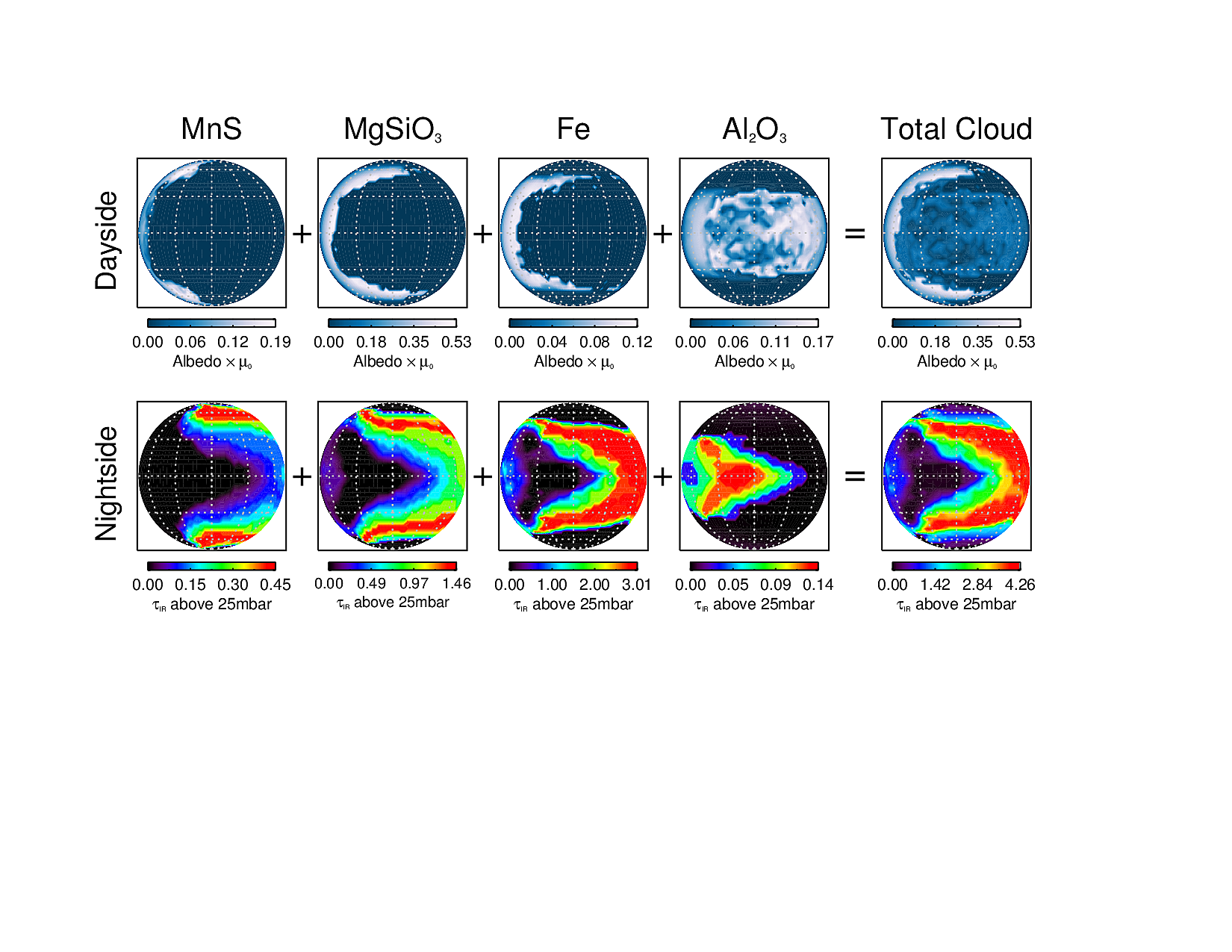}
\caption{(top) Contribution of each cloud species to the total dayside reflectance, assuming weak vertical mixing limits clouds to roughly a scale height in thickness, temperature permitting, and 10\% condensation efficiency. Dayside atmospheric albedos are multiplied by the cosine of the stellar incidence angles to appropriately account for the geometric effect of diffuse reflection from a sphere, as approximated in our two-stream calculations. Clouds are assumed to be composed of small particle (0.2 $\mu$m). MnS, MgSiO${_3}$, and iron clouds form preferentially near the cooler wester terminator and high latitudes; the warmer condensation curve of Al${_2}$O${_3}$ allows thin clouds to form over much of the dayside, contributing to the total albedo.  The poles have low albedo as the clouds form and terminate deeper in colder regions. (bottom) Likewise, the IR optical thickness of aerosols on the nightside, integrated from the top of the atmosphere down to the 25 mbar level, roughly corresponding to the pressure of the clear atmosphere infrared photosphere. A majority of the thermal cloud opacity is due to iron condensate due to its relatively large thermal extinction efficiency. Over this pressure range, the thickest clouds form at high latitudes and eastward of the anti-stellar point. A thin cloud of Al${_2}$O${_3}$ forms at warmer temperatures to the west.}
\label{fig:compactcom}
\end{figure*}
 
\begin{figure*}[th!]
\centering
\includegraphics[clip,trim=1.5in 5.5in 3.2in 2in, width=\textwidth]{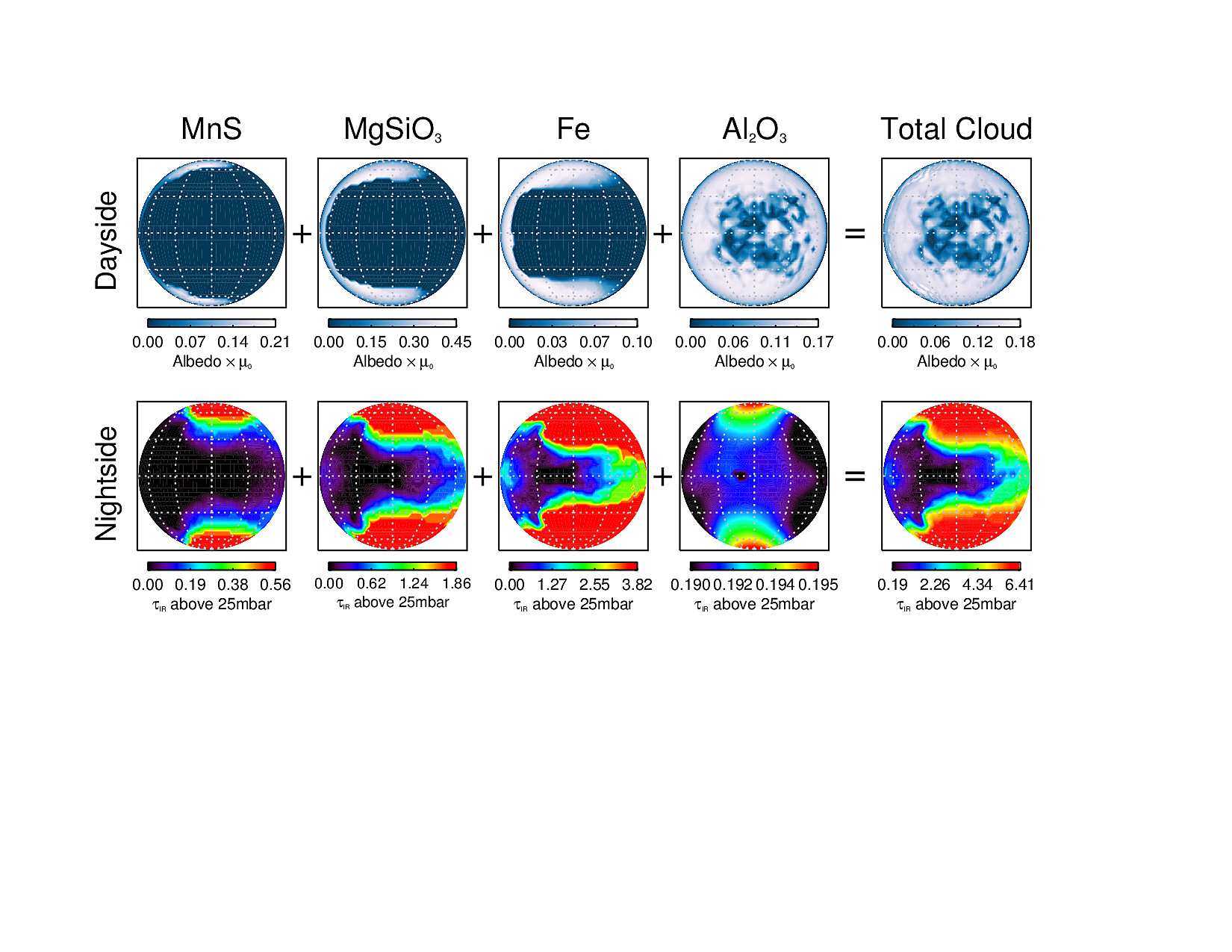}
\caption{As in Figure \ref{fig:compactcom}, but for clouds with vertical thicknesses permitted to extend up to 100 $\mu$bar, temperature permitting, to mimic strong vertical mixing. Compared to the scale-height cloud cases, the extended clouds produce a more uniform reflectance and opacity in the observable atmosphere as the cloud base pressures are less consequential, though their general distributions are qualitatively similar, with the exception of Al${_2}$O${_3}$, which now covers most of the planet in a thin cloud at observable pressures. With all species clouds mixing up to 100 microbar, the lower single scattering albedos of the Al${_2}$O${_3}$, MnS, and Fe clouds dilute the albedo of the MgSiO${_3}$ cloud in the total mixture.  The broken cloud coverage due to radiative feedback is apparent on the dayside.}
\label{fig:extendedcom}
\end{figure*}

To understand the role of each cloud type, we plotted the individual contribution of each species for both the compact and extended clouds models, focusing on just the thicker cloud cases. Given differences in the condensation curves for each of the included potential condensates, each cloud had a unique distribution depending on the temperature field. MgSiO${_3}$, MnS, and Fe have cooler condensation curves, and thus they preferentially form in regions of cooler temperatures along the western terminator and northern latitudes. The relatively higher condensation curve for Al${_2}$O${_3}$ allows clouds to form at warmer temperatures over most of the planet, including most of the dayside.  We found that the precise locations of clouds and their observability depends on the assumed particle properties and whether or not radiative feedback is included. 

To illustrate the differences due to vertical distribution and effective particle properties of each cloud, we computed the dayside reflection and nightside attenuating optical thickness for the total cloud mixture and each component, as shown in Figures \ref{fig:compactcom}-\ref{fig:extendedcom}. 

The dayside reflectances are expressed as the local top-of-the-atmosphere albedo multiplied by the cosine of the incidence angle ${\mu_0}$ (or equivalently, the cosine of the emission); the correction by ${\mu_0}$ is intended to account for reduced contribution approaching the limbs of the spherical planet given the sub-stellar point is at center.  

The abundance, extinction efficiency, single scattering albedo, and asymmetry parameter all contribute to the intrinsic albedo of the cloud, and each of these quantities differ between cloud types. In terms of potential to form brightly reflective clouds, the MgSiO${_3}$ cloud has the greatest abundance and a very high single scattering albedo and thus the potential to contribute considerably to the total albedo. Iron is the second most abundant, but its low single scattering albedo ($\sim$0.7) reduces its potential reflectivity, particularly for smaller particle sizes. Al${_2}$O${_3}$ is an order of magnitude less abundant and has a low albedo, so less intrinsically reflective. MnS has yet another order of magnitude less abundance, but it is a highly conservative scatterer, and so it could potentially significantly contribute to the total albedo. In all cases, particles of each cloud were small, so asymmetry parameters were low and the clouds could backscatter efficiently.  If particles happen to be larger, they would strongly forward scattered, reducing their contribution to the observable albedo, but modeling suggests particles near the top of atmosphere are likely sub-micron in radius \citep{Parmentier2013, Lee_dynamical_clouds2017, Lines2018}.

The vertical positioning is perhaps even more important.  Higher altitude clouds can potentially contribute more to the local albedo, as they are above more of the absorbing atmosphere.  As a cloud's top moved deeper into the absorbing atmosphere, its ability to contribute to the observed reflectance diminished.  Likewise, the relative mixing or layering of clouds was very significant.   In our compact cloud case, the MgSiO${_3}$ cloud formed at a greater height than the intrinsically less reflective Fe and Al${_2}$O${_3}$ clouds; the latter clouds formed deeper and terminated beneath the MgSiO${_3}$. As a result, the local albedo along the cool western terminator and norther latitudes was dominated by the overlying MgSiO${_3}$ reflectance. In contrast, if the atmosphere is well mixed, Al${_2}$O${_3}$ and Fe extended to the top of the model and mixed with the MgSiO${_3}$ to reduce the net single scattering albedo of the total cloud mixture, as can be seen comparing Figures \ref{fig:compactcom} and \ref{fig:extendedcom}.  This of course assumes the total scattering properties may be approximated by a weighted mixture; chemistry and cloud microphysics could result in one species coating another, yielding properties more like the outer most layer, in which case the computed albedos may differ.

On the shadowed nightside, where reflectance is unimportant, we present cloud thermal opacity above 25 mbar (roughly the pressure of the clear atmosphere's thermal photosphere).  This picture is simpler since the total opacity is essentially determined by the sum of the component cloud infrared opacities.  A majority of the aerosol thermal opacity was due to iron condensate, given its moderate abundance and relatively large thermal extinction efficiency, followed by MgSiO${_3}$.  Most of the attenuating cloud opacity was located eastward and poleward of the anti-stellar point, although, the compact clouds attenuated less outgoing thermal radiation at the poles simply because clouds formed and extinguished at deeper pressures.   Likewise, Al${_2}$O${_3}$ produced an optically thin cloud west of anti-stellar in the compact case, but formed a more uniform cloud covering all the nightside when permitted to extend to the top of the model. 

\subsection{Predicted Reflectance and Emission}

To evaluate the observable implications of our different models, we computed the dayside top-of-the-atmosphere albedos, out-going thermal emissions, and resulting visible and thermal phase curves for each using our two different approaches to processing clouds.

\subsubsection{Total Dayside Albedos} \label{albedos}

\begin{figure*}[th!]
\includegraphics[clip, trim=0in 3.1in 0in 7.9in,width=\textwidth]{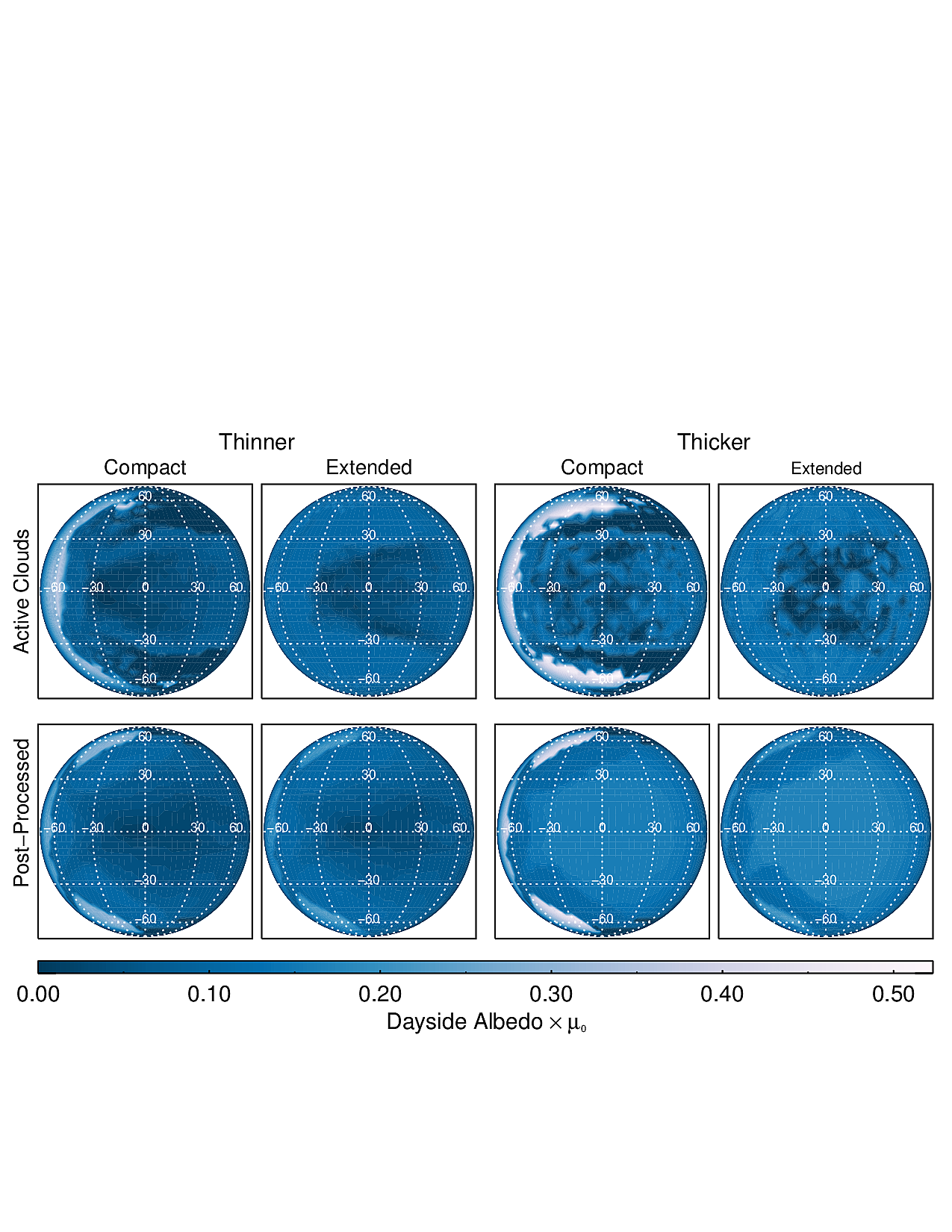}
\caption{Reflectance, expressed as top-of-the-atmosphere albedos multiplied by the cosine of the emission angle, for the compact-thin cloud (left), extended-thin cloud (middle-lift), and compact-thick cloud (middle-right), and extended-thick cloud (right) models, using active (top) and post-processed (bottom) approaches. The compact cloud models allows reflective MnS and MgSiO${_3}$ clouds to reach above less reflective Fe and Al${_2}$O${_3}$ clouds, while the extended clouds mix all the cloud species together; in the thicker case, the mixing of cloud species reduces the albedo, and radiative feedback causes cloud cover to be broken on the dayside in the active cloud modeling.}
\label{fig:albedo}
\end{figure*}

The reflectance from the planet was dependent on the cloud distribution, abundance, and particle scattering properties as discussed in Section 3.3. As Figure \ref{fig:albedo} shows, the two models assuming greater optical thicknesses were unsurprisingly more reflective. Of the two thicker models, the compact was the most reflective, despite the limited extent. This was because the bright, high altitude MnS and MgSiO${_3}$ clouds reached above less reflective Fe and Al${_2}$O${_3}$ clouds; if the clouds were extended, the mixed cloud resulted in a lower albedo despite the greater overall cloud cover. If the clouds were post-processed onto the clear temperature field, the compact cloud case had a global spherical albedo of 0.264, while the extended cloud had only a slightly lower value of 0.257.  When clouds were actively included, the thick compact cloud had a spherical albedo of 0.230, while the extend had an albedo of 0.202. In these latter cases, the albedos were lower and the differences were greater because the radiative feedbacks vaporized clouds on the dayside, resulting in a less uniform coverage.

The trend was opposite for the thinner clouds, as the global reflectance was less influenced by the relative single scattering albedo of different cloud species, and more influenced by the overall spatial coverage.  The active thin cases showed far less broken coverage than their thicker counterparts, and the difference between the extended case's albedo (0.173) and the compact case's (0.153) was less. The post-processed models again show higher albedos in each case, with spherical albedos 0.180 and 0.166 for extended and compact cases, respectively.  The relatively greater reflectance for all the post-processed cases shows the role of feedback in limiting the cloud cover in our models; however, it is important to note that different scattering properties would yield different results.  Our initial tests showed that if only highly conservative scattering clouds are included (i.e. ${\varpi_0}$ $\sim$1), a positive radiative feedback causes the atmosphere to grow cold and cloudier, quickly clouding over completely. So the result of the feedback depends on the nature of the scatters.

Calculations for the atmospheric albedos neglected Rayleigh scattering, which could significantly increase the albedo of the clear or thinly clouded atmosphere.  The potential contribution from Rayleigh scattering depends on the precise composition of the atmosphere and the wavelengths at which it is observed.  We previously determined that a spherical albedo of 0.1 at 400 nm was possible in our simple two-stream framework due to Rayleigh scattering alone \citep{Roman&Rauscher2017}, while more rigorous modeling by \cite{Demory2013} and \cite{MnI2015} suggested spherical albedos of 0.2 or more could be possible depending on the amount of atmospheric absorbers.  \cite{gao2017sulfur} noted that global sulfur photochemical hazes could strongly absorb shortward of 0.4 $\mu$m, resulting in albedos  $\textless$ 0.1, in contrast to the higher albedos from scattering in a clear atmosphere. In any case, additional Rayleigh scattering would increase the albedo and reduce inhomogeneity in the reflectance.

We note that all of our calculations yield a spherical albedo that is a factor of two less than the spherical albedo of Kepler-7b over the \textit{Kepler} passband, which was estimated to be in the range 0.4–0.5 by \cite{MnI2015} (or, alternatively, a geometric albedo of 0.35 $\pm$ 0.02 \citep{Demory2013}). In general, achieving an albedo this great would require a large abundance of highly reflective scatterers at sufficiently high altitudes above the absorbing atmosphere. This can be quite easily achieved by placing just a few optical depths of aerosols uniformly over the disk \citep{Roman&Rauscher2017}, as one might expect from a photochemical haze; however, for an inhomogeneous condensate cloud, this requires far greater aerosol optical thicknesses ($\sim$ 150 or greater above 10 mbar) depending on the precise coverage, as calculated by \cite{MnI2015} and tested in a GCM by \cite{Roman&Rauscher2017}.  Our present simulations permit comparable aerosol optical thicknesses, but the observed reflectance of Kepler-7b does not fall out of the model naturally under the assumed conditions.  Given that we limit the potential optical thicknesses to 10\% of the theoretical maximum values (by our estimate), it is tempting to solve the albedo deficiency by simply adding more aerosol mass, but our initial tests suffer from numerical instabilities at higher abundances due to strong absorption, as discussed in Sec. \ref{scatparamdisc}.  Furthermore, if the particle only scatter (i.e. $\varpi$$_0$ $\sim$ 1), positive radiative feedbacks can easily cause the planet to become too cold, globally cloudy, and far too reflective. This shows a limitation of our model, which lacks physics of cloud limiting processes (such as rainout and advection), but it also displays the sensitive dependence on the combination of cloud species and scattering parameters used in the model, which may require some balancing to self-consistently achieve the observed values of Kepler-7b.

\subsubsection{Global Thermal Emissions and Energy Balance}

 \begin{figure*}[th!]
\includegraphics[clip, trim=0in 0in 0in 0in,width=\textwidth]{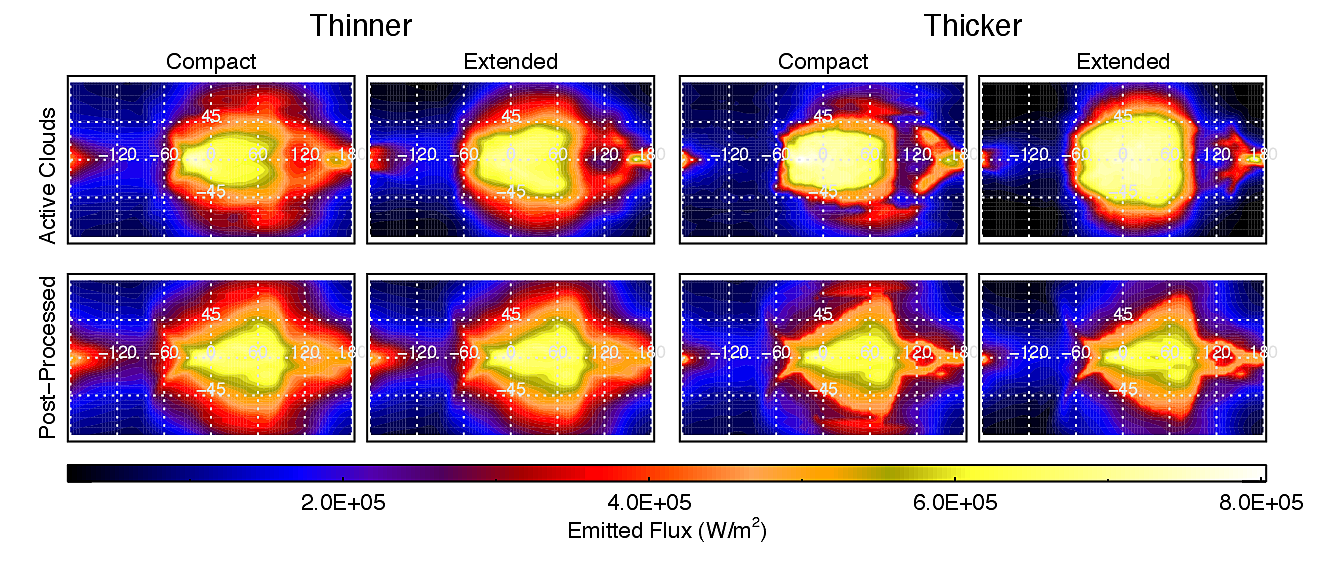}
\caption{Mapped top-of-the-atmosphere thermal emission from the compact-thin cloud (left), extended-thin cloud (middle-lift), and compact-thick cloud (middle-right), and extended-thick cloud (right) models, using active (top) and post-processed (bottom) approaches. The substellar point is at the center (0$^{\circ}$, 0$^{\circ}$). As clouds reduce emission on the nightside and towards the poles, dayside emission increases in the active cloud models, in contrast to the post-processed models.}
\label{fig:emission}
\end{figure*}

The total thermal emission from the top of each modeled atmosphere are shown in Figure \ref{fig:emission}.  In all cases, clouds reduced the emission relative to the clear atmosphere over the nightside, in contrast to what we had found using static, thermally insulating clouds \citep{Roman&Rauscher2017}, but the choice of cloud model and processing significantly affected the pattern of emission. Recall that our method of post-processing simply added clouds to the final iteration of the clear simulation, and so they could affect the emission only by attenuating outgoing radiation. As such, the patterns for the post-processed cases were that of the clear atmosphere with fluxes reduced in locations where clouds existed above the clear infrared photosphere. The differences are relatively subtle and they appear qualitatively similar to each other, with globally reduced emission and an eastward shift of the maximum, consistent with standard hot Jupiter patterns \citep{showman2009atmospheric,Dobbs-Dixon2010,RauscherMenou2010,Heng2011,Mayne2014}.  The poles in the post-processed compact case emitted more than the other post-processed cases since those clouds were confined beneath the height of maximum emission to space. 

In contrast, the active cloud modeling allowed the atmosphere to respond and adjust to the cloud scattering, and so the actual heating pattern differed. This resulted in more strongly reduced nightside emission and enhanced dayside emission relative to the clear case.  The intense dayside emission was in part due to the cloud heating and the partial clearing caused by the intense instellation on the dayside, but it was fundamentally a matter of global energy balance. As less energy radiatively escaped from the mostly clouded nightside, more was forced to emerge from the dayside in order to maintain a global quasi-equilibrium.  The post-processed solutions were not subject to this rough energy balance and thus did not see an increase in dayside emission or an overall reduction in energy due to higher spherical albedos.  In the post-processed thin cloud cases, the total outgoing energy was determined by the instantaneous reflectance from clouds plus the atmosphere's intrinsic thermal emission, partly reduced by the presence of attenuating clouds, just as it was in the active cases; however, unlike the active cases, the intrinsic thermal emission in the post-processed cases was set by the temperature of the \emph{clear} atmosphere, which is warmer due to its lower albedo. The post-processed cases therefore intrinsically emit too much thermal radiation. In the thin cloud cases, the total outgoing radiation exceeded the incoming radiation by 10\% when we assumed an extended cloud, and by 12\% when using the compact cloud. The imbalance appears somewhat less for the thicker cloud cases (2\% and 8\% for the extended and compact cases, respectively), but this was simply due to the thicker clouds' ability to block more of the inflated intrinsic emission.

It is worth noting that though energy is balanced in our active models, the applied double gray approximation limits the accuracy of the computed transmissions, as discussed in Section \ref{2xgraydisc}.  In particular, the emission is likely underestimated in the coldest, cloudiest regions, and to maintain energy balance, overestimated in the warm clear regions.  A direct comparison between similar modeling using double-gray and non-gray should be the focus of future work.
  
\subsubsection{Reflected and Thermal Phase Curves}

\begin{figure*}[th!]
\centering
\includegraphics[clip, trim=0in .5in 0.2in 5.4in, width=\textwidth]{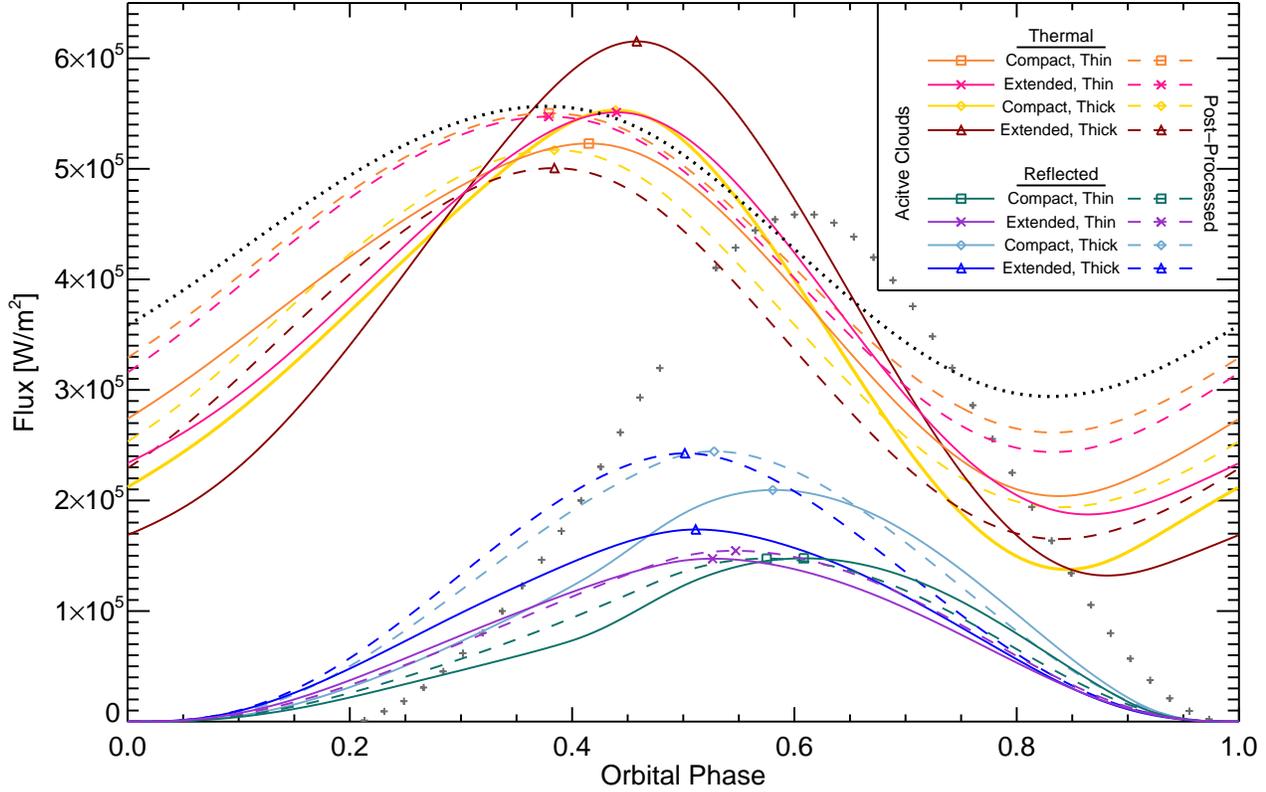}
\caption{Phase curves for different cloud models computed from simulations with active clouds (solid) and post-processed clouds (dashed), along with the clear model (black dotted).  Orbital phases of zero and one correspond to the when the center of the nightside (i.e. antistellar point) faces the observer, while 0.5 corresponds to the substellar point. The curves are color coded, with the location of the maximum emission and reflectance indicated, as shown in the key. The observationally inferred reflected-light phase curve \citep{[\textit{Kepler} bandpass][]Demory2013} is represented by the gray + symbols. The active cloud model results have greater amplitude thermal phase curves with lesser phase shifts (less eastward) compared to the post-processed and clear model results. In reflected light, the phase curves for the active clouds and post-processed cases are similar in amplitude and phasing for the thin clouds.  For the thicker clouds, both post-processed curves have greater amplitude, but slightly lesser phase shift compared to the active clouds.}
\label{fig:activevsppphase}
\end{figure*}

The reflected light and thermal emission were used to compute the reflected and thermal phase curves shown in Figure \ref{fig:activevsppphase}.  The figure shows the curves for each case of cloud properties using the active and post-processing approaches.  For emission curves, the post-processed curves were a relatively similar to each other, as would be expected from the discussion of emission maps above; the clouds simply attenuated emission without altering the temperature field, and so their ability to dramatically alter the pattern of emission and differentiate from one another was limited. The thermal phase curves clearly showed that these post-processed solutions are essentially identical in phasing to the clear atmosphere curve, with the peaks shifted $\sim$ 41$^{\circ}$- 45$^{\circ}$ to the east of sub-stellar, though with significantly reduced flux, particularly on the nightside.  The flux is preferentially reduced on the nightside where clouds are thicker, modestly increasing the amplitude of the curve. 

The active cloud models, however, show how the radiative feedbacks can significantly increase the amplitude and alter the phasing of the emission curve.  The reduced nightside flux and compensating greater dayside flux resulted in greater amplitudes than we saw for the clear and post-processed curves. The extended cloud emission curves had larger amplitudes than the corresponding compact cloud cases as a result of the higher clouds penetrating farther into the thermal photosphere. Interestingly, in all four active cloud cases, the phasing is such that the hotspot was closer to the substellar point.  The eastward shifts were relatively modest, falling between $\sim$ 14$^{\circ}$ and 29$^{\circ}$ east of substellar, decreasing with increasing amplitude. The clear and post-processed curves showed an offset of $\sim$43$^{\circ}$ from substellar. 

In reflected light phase curves, the post-processed cases show peaks and shifts up to 25\% greater than the active cloud cases for the thicker cloud cases, consistent with the radiative feedback reducing dayside cloud coverage in the active models. For the thinner clouds, the differences between the two approaches were less than 10\%.

The phasing of the reflected phase curve peak was determined by the distribution of the reflective clouds. In the compact active cases, the exposed, bright, western-terminator MgSiO${_3}$ and MnS clouds produced offsets of roughly 29$^{\circ}$ to 40$^{\circ}$. When clouds were extended, this shift was partly offset by lower albedo clouds that covered much of the dayside, falling at less than 20$^{\circ}$. 

It is interesting to compare these reflected light phase curves to those observed and previously modeled for Kepler-7b, whose planet parameters we have adopted for this study.  \cite{MnI2015} found that brightly reflecting, thick clouds of small particles located west of substellar could were needed to best fit the asymmetry in the reflected phase curve (offset westward $\sim$ 41 $\pm$ 12$^{\circ}$) as observed by \cite{Demory2013}. \cite{Webber2015} came to a similar conclusion but found that their models fell short of reproducing the observations in both albedo and phase shift. Presently, by letting the cloud cover and consequent reflectance be determined self-consistently by the temperature field, we find we can only match the phase shift in some cases (the compact cases) and always fall short of the observed amplitude by more than a factor of two. While it is not surprising that our albedos are too low (as discussed in Section \ref{albedos}) given that we only include 10\% of the potential condensible mass for just a few of the probable cloud species (two of which have relatively low single scattering albedos), increasing the cloud abundances to match the albedo does not easily improve the fit to the phase curve.  This is because a higher albedo will result in a colder atmosphere that can form clouds globally, producing little asymmetry in the reflectance, as is seen in the cloudy simulations of \cite{Lines2018}.  Our preliminary tests show similar effects, but more work is needed to understand the role of neglected processes and omitted cloud species on the reflected phase curves. It is conceivable that, given a diversity of aerosols, a strongly asymmetric reflectance in a higher albedo atmosphere can be maintained if conservative scattering and cooling along the western terminator is balanced by aerosol absorption and heating in clouds to the east.

Finally, we contrast these results with our previous modeling, which neglected any radiative feedback.  In that previous work,  very high, thick clouds were fixed in place to reproduce reflected phase curves, while the infrared infrared phase curves were then predicted. We had found eastward shifts of $\sim$43$^{\circ}$-68$^{\circ}$ \textemdash greater than expected for the clear atmosphere alone.  Now, with a more complex, physically motivated treatment of multiple clouds that each respond to changes in the temperature structure, we find our model predicts infrared phase curves that are less than expected for a clear atmosphere. This difference can largely be explained by the role of aerosol radiative feedback in context of our double-gray model.  Given the simple temperature dependence simulated, aerosols in our model grow thicker where it is colder, and become thinner were it is hot.  On the dayside, the intense heating of optical absorbing aerosols in the substellar region causes local heating, which in turn results in patchy cloud coverage as clouds continually evaporate and reform. This creates a window through the clouds that allows for more radiation to penetrate.  However, critical to the infrared phase curves, it also creates a more transparent window through which more thermal radiation can escape from the otherwise cloudy, opaque atmosphere. So the thermal response to heating in the visible channel sets the pattern for how heat escapes in the thermal channel.  Since the stellar heating is centered at the substellar point, the resulting phase curves peak closer to the substellar point, and hence have less phase offset. Without radiative feedback, this would not occur.  Together, our present and previous models demonstrate how feedback can alter the conclusions drawn from GCM modeling.

\section{Conclusions} \label{conclusions}

We modeled a hot Jupiter atmosphere using a GCM with a simple temperature-dependent cloud scheme that includes basic radiative feedbacks.  We ran simulations for four cases, each with a different assumption regarding the vertical distribution and cloud opacities.  For each case, we processed clouds using two different approaches. In our active-cloud approach, clouds were included throughout the duration of the model and allowed to alter and respond to the temperature field; in our post-processing approach, clouds were only passively added to the final results of a clear atmospheric model.  This comparison highlighted the important role of a simple feedback mechanism and how assumptions regarding cloud properties may affect results.  We draw the following conclusions:

\begin{itemize}
  \item Predicted cloud distributions differ depending on whether radiative feedback processes are included or not. Simulations including active radiative feedback resulted in less cloud cover along the equator and more at high latitudes than that found in results post-processed from clear models. Differences were greatest for thick and extended clouds, though all our tested cloud models produced significant differences in cloud cover, temperature, thermal emission, and phase curves. In general, differences in phase curve amplitudes and shifts were roughly 10\% or less for the thinner clouds (visible optical thickness per bar of $\sim$1,400) and as much as 25\% for the thicker clouds ($\sim$ 14,000 $\tau$ per bar).
    \item Radiative feedback from clouds can significantly alter the temperature field, depending primarily on aerosol scattering properties and abundances.  Highly reflective and abundant MgSiO${_3}$ and MnS clouds can significantly cool the underlying atmosphere and increase the planetary albedo.  Absorbing Al${_2}$O${_3}$ and Fe can heat the atmosphere.  In particular, Al${_2}$O${_3}$ was found to raise the substellar temperatures and force them towards the Al${_2}$O${_3}$ condensation curve as the mineral fluctuates between evaporating and condensing. Other condensates may behave similarly in atmospheres of different temperatures. Processes neglected in this modeling, such as rain out, latent heating, and advection of cloud particles, may modify these effects. 
    \item The vertical positioning of clouds is significant.  Clouds forming higher in the atmosphere have a stronger influence on the observed reflectance and emission.  If clouds are vertically layered, the properties of the highest cloud can dominate. If clouds are vertically well mixed, darker component particles may easily reduce the overall albedo of the mixture depending on the cloud chemistry and microphysics. 
    \item When clouds are post-processed, the global energy balance is typically not conserved because the atmosphere cannot thermally adjust to the altered fluxes. This was true for our thicker and thinner cloud models. On the dayside, this means the increased reflectance will not necessarily match the diminished emission.  On the nightside, clouds may reduce emission from the underlying atmosphere without a compensating increase in emission elsewhere.  When radiative feedback is actively included throughout the simulation, the global energy balance is conserved.  In these cases, an increased dayside albedo can reduce the global emission, and an attenuated emission on a cloudy nightside will be compensated by an increase in emission on a clearer dayside, creating a greater day-night contrast.
     \item Our computed thermal phase curves from post-processed simulations were similar in phase shift and amplitude to each other, and similar to the clear atmosphere results, but with reduced flux. If radiative feedback was included, in all cases the amplitude of the curves significantly increased while the eastward phase shift decreased due to more radiation escaping through clearings in the substellar region; however, the amplitude of the predicted phase curve and day-night contrast may be overestimated by the double-gray approximation.
      \item Compact cloud models with active radiative feedback produced reflected light phase curves with the greatest phase shifts, consistent to that inferred for Kepler-7b, though the amplitudes were deficient by a factor of two or more in all cases. This suggest that even higher and/or thicker reflective clouds along western limb are required to match observations, possibly in combination with warming, less reflective clouds to the east. The highest amplitude curves were produced by thick, post-processed clouds.      

 \end{itemize}     
 
Though our modeling neglects or simplifies much of the complex aerosol physics, it neatly isolates the role of scattering and radiative cloud feedbacks for a typical hot Jupiter atmospheres.  The basic conclusions illustrate how aerosol scattering may alter the temperature field and consequent observations.  As the radiative effects associated with clouds alter the local atmospheric temperatures and fluxes, the greater atmosphere responds to fundamentally preserve the global energy balance. Existing cloud distributions will naturally change to respond to the altered temperature field until a quasi-steady state equilibrium is reached. Instantaneous post-processing does not allow for these adjustments, and so the predicted cloud distribution, emission, and reflectance will differ. While post-processing clear simulations can be appropriate for determining basic characterizations of cloud distributions in an atmosphere, it neglects the important response of the atmosphere that in turn affects the cloud distribution.  Care should be taken in using post-processing when clouds are thick enough to provide significant scattering, which may be common given the expected abundances of condensible gases in solar-composition atmospheres. 

\acknowledgements 
This research was supported by NASA Astrophysics Theory Program grant NNX17AG25G. 

\bibliography{mybib} 

\begin{thebibliography}{}
\expandafter\ifx\csname natexlab\endcsname\relax\def\natexlab#1{#1}\fi

\bibitem[{{Ackerman} \& {Marley}(2001)}]{Ackerman&Marley1989}
{Ackerman}, A.~S., \& {Marley}, M.~S. 2001, \apj, 556, 872

\bibitem[{{Ag{\'u}ndez} {et~al.}(2014){Ag{\'u}ndez}, {Parmentier}, {Venot},
  {Hersant}, \& {Selsis}}]{Agundez2014}
{Ag{\'u}ndez}, M., {Parmentier}, V., {Venot}, O., {Hersant}, F., \& {Selsis},
  F. 2014, \aap, 564, A73

\bibitem[{{Banfield} {et~al.}(1998){Banfield}, {Conrath}, {Gierasch},
  {Nicholson}, \& {Matthews}}]{Banfield1998a}
{Banfield}, D., {Conrath}, B.~J., {Gierasch}, P.~J., {Nicholson}, P.~D., \&
  {Matthews}, K. 1998, \icarus, 134, 11

\bibitem[{Bauer {et~al.}(2015)Bauer, Thorpe, \&
  Brunet}]{bauer2015numericalmodels}
Bauer, P., Thorpe, A., \& Brunet, G. 2015, Nature, 525, 47

\bibitem[{Bohren \& Albrecht(2000)}]{bohren2000atmospheric}
Bohren, C.~F., \& Albrecht, B.~A. 2000, Atmospheric thermodynamics

\bibitem[{{Burrows} \& {Sharp}(1999)}]{Burrows&Sharp1999}
{Burrows}, A., \& {Sharp}, C.~M. 1999, \apj, 512, 843

\bibitem[{{de Rooij} \& {van der Stap}(1984)}]{deRooij1984}
{de Rooij}, W.~A., \& {van der Stap}, C.~C.~A.~H. 1984, \aap, 131, 237

\bibitem[{{Demory} {et~al.}(2013){Demory}, {de Wit}, {Lewis}, {Fortney},
  {Zsom}, {Seager}, {Knutson}, {Heng}, {Madhusudhan}, {Gillon}, {Barclay},
  {Desert}, {Parmentier}, \& {Cowan}}]{Demory2013}
{Demory}, B.-O., {de Wit}, J., {Lewis}, N., {et~al.} 2013, \apjl, 776, L25

\bibitem[{{Dobbs-Dixon} {et~al.}(2010){Dobbs-Dixon}, {Cumming}, \&
  {Lin}}]{Dobbs-Dixon2010}
{Dobbs-Dixon}, I., {Cumming}, A., \& {Lin}, D.~N.~C. 2010, \apj, 710, 1395

\bibitem[{{Esteves} {et~al.}(2015){Esteves}, {De Mooij}, \&
  {Jayawardhana}}]{Esteves2015phasecurves}
{Esteves}, L.~J., {De Mooij}, E.~J.~W., \& {Jayawardhana}, R. 2015, \apj, 804,
  150

\bibitem[{Fortney {et~al.}(2008)Fortney, Lodders, Marley, \&
  Freedman}]{Fortney2008unified}
Fortney, J.~J., Lodders, K., Marley, M.~S., \& Freedman, R.~S. 2008, The
  Astrophysical Journal, 678, 1419

\bibitem[{Gao(2017)}]{Gao2017clouds}
Gao, P. 2017, PhD thesis, California Institute of Technology

\bibitem[{{Gao} {et~al.}(2018){Gao}, {Marley}, \&
  {Ackerman}}]{GaoMarley&Ackerman2018}
{Gao}, P., {Marley}, M.~S., \& {Ackerman}, A.~S. 2018, \apj, 855, 86

\bibitem[{Gao {et~al.}(2017)Gao, Marley, Zahnle, Robinson, \&
  Lewis}]{gao2017sulfur}
Gao, P., Marley, M.~S., Zahnle, K., Robinson, T.~D., \& Lewis, N.~K. 2017, The
  Astronomical Journal, 153, 139

\bibitem[{{Gibson} {et~al.}(2013){Gibson}, {Aigrain}, {Barstow}, {Evans},
  {Fletcher}, \& {Irwin}}]{Gibson2013}
{Gibson}, N.~P., {Aigrain}, S., {Barstow}, J.~K., {et~al.} 2013, \mnras, 428,
  3680

\bibitem[{{Gibson} {et~al.}(2012){Gibson}, {Aigrain}, {Pont}, {Sing},
  {D{\'e}sert}, {Evans}, {Henry}, {Husnoo}, \& {Knutson}}]{Gibson2012}
{Gibson}, N.~P., {Aigrain}, S., {Pont}, F., {et~al.} 2012, \mnras, 422, 753

\bibitem[{Guillot(2010)}]{guillot2010radiative}
Guillot, T. 2010, Astronomy \& Astrophysics, 520, A27

\bibitem[{Hansen \& Travis(1974)}]{hansen_travis_1974}
Hansen, J.~E., \& Travis, L.~D. 1974, Space science reviews, 16, 527

\bibitem[{{Heng} \& {Demory}(2013)}]{HengDemory2013}
{Heng}, K., \& {Demory}, B.-O. 2013, \apj, 777, 100

\bibitem[{{Heng} {et~al.}(2011){Heng}, {Menou}, \& {Phillipps}}]{Heng2011}
{Heng}, K., {Menou}, K., \& {Phillipps}, P.~J. 2011, \mnras, 413, 2380

\bibitem[{Hoskins \& Simmons(1975)}]{Hoskins1975}
Hoskins, B., \& Simmons, A. 1975, Quarterly Journal of the Royal Meteorological
  Society, 101, 637

\bibitem[{{Hu} {et~al.}(2015){Hu}, {Demory}, {Seager}, {Lewis}, \&
  {Showman}}]{Hu2015}
{Hu}, R., {Demory}, B.-O., {Seager}, S., {Lewis}, N., \& {Showman}, A.~P. 2015,
  \apj, 802, 51

\bibitem[{{Kataria} {et~al.}(2015){Kataria}, {Showman}, {Fortney}, {Stevenson},
  {Line}, {Kreidberg}, {Bean}, \& {D{\'e}sert}}]{Kataria2015}
{Kataria}, T., {Showman}, A.~P., {Fortney}, J.~J., {et~al.} 2015, \apj, 801, 86

\bibitem[{{Kataria} {et~al.}(2016){Kataria}, {Sing}, {Lewis}, {Visscher},
  {Showman}, {Fortney}, \& {Marley}}]{Kataria2016}
{Kataria}, T., {Sing}, D.~K., {Lewis}, N.~K., {et~al.} 2016, \apj, 821, 9

\bibitem[{{Kitzmann} \& {Heng}(2018)}]{KitzmannHeng2018}
{Kitzmann}, D., \& {Heng}, K. 2018, \mnras, 475, 94

\bibitem[{Kreidberg {et~al.}(2018)Kreidberg, Line, Thorngren, Morley, \&
  Stevenson}]{Kreidberg2018}
Kreidberg, L., Line, M.~R., Thorngren, D., Morley, C.~V., \& Stevenson, K.~B.
  2018, The Astrophysical Journal Letters, 858, L6

\bibitem[{{Latham} {et~al.}(2010){Latham}, {Borucki}, {Koch}, {Brown},
  {Buchhave}, {Basri}, {Batalha}, {Caldwell}, {Cochran}, {Dunham}, {F{\H
  u}r{\'e}sz}, {Gautier}, {Geary}, {Gilliland}, {Howell}, {Jenkins},
  {Lissauer}, {Marcy}, {Monet}, {Rowe}, \& {Sasselov}}]{Latham2010}
{Latham}, D.~W., {Borucki}, W.~J., {Koch}, D.~G., {et~al.} 2010, \apjl, 713,
  L140

\bibitem[{{Lee} {et~al.}(2016){Lee}, {Dobbs-Dixon}, {Helling}, {Bognar}, \&
  {Woitke}}]{Lee_kitchensink2016}
{Lee}, G., {Dobbs-Dixon}, I., {Helling}, C., {Bognar}, K., \& {Woitke}, P.
  2016, \aap, 594, A48

\bibitem[{{Lee} {et~al.}(2017){Lee}, {Wood}, {Dobbs-Dixon}, {Rice}, \&
  {Helling}}]{Lee_dynamical_clouds2017}
{Lee}, G.~K.~H., {Wood}, K., {Dobbs-Dixon}, I., {Rice}, A., \& {Helling}, C.
  2017, \aap, 601, A22

\bibitem[{{Lewis} {et~al.}(2017){Lewis}, {Parmentier}, {Kataria}, {de Wit},
  {Showman}, {Fortney}, \& {Marley}}]{lewis2017cloudevo}
{Lewis}, N.~K., {Parmentier}, V., {Kataria}, T., {et~al.} 2017, ArXiv e-prints,
  arXiv:1706.00466

\bibitem[{{Lines} {et~al.}(2018){Lines}, {Mayne}, {Boutle}, {Manners}, {Lee},
  {Helling}, {Drummond}, {Amundsen}, {Goyal}, {Acreman}, {Tremblin}, \&
  {Kerslake}}]{Lines2018}
{Lines}, S., {Mayne}, N.~J., {Boutle}, I.~A., {et~al.} 2018, ArXiv e-prints,
  arXiv:1803.00226

\bibitem[{{Lodders}(2010)}]{Lodders2010}
{Lodders}, K. 2010, {Exoplanet Chemistry}, ed. R.~{Barnes}, 157

\bibitem[{L{\'o}pez(1977)}]{lopez_1977lognormal}
L{\'o}pez, R.~E. 1977, Monthly Weather Review, 105, 865

\bibitem[{{Marley} {et~al.}(2013){Marley}, {Ackerman}, {Cuzzi}, \&
  {Kitzmann}}]{Marley2013exoclouds}
{Marley}, M.~S., {Ackerman}, A.~S., {Cuzzi}, J.~N., \& {Kitzmann}, D. 2013,
  {Clouds and Hazes in Exoplanet Atmospheres}, ed. S.~J. {Mackwell}, A.~A.
  {Simon-Miller}, J.~W. {Harder}, \& M.~A. {Bullock}, 367--391

\bibitem[{{Mayne} {et~al.}(2014){Mayne}, {Baraffe}, {Acreman}, {Smith},
  {Browning}, {Sk{\aa}lid Amundsen}, {Wood}, {Thuburn}, \&
  {Jackson}}]{Mayne2014}
{Mayne}, N.~J., {Baraffe}, I., {Acreman}, D.~M., {et~al.} 2014, \aap, 561, A1

\bibitem[{{Mbarek} \& {Kempton}(2016)}]{Mbarek&Kempton2016}
{Mbarek}, R., \& {Kempton}, E.~M.-R. 2016, \apj, 827, 121

\bibitem[{{Mendon{\c c}a} {et~al.}(2018){Mendon{\c c}a}, {Malik}, {Demory}, \&
  {Heng}}]{Mendoca2018}
{Mendon{\c c}a}, J.~M., {Malik}, M., {Demory}, B.-O., \& {Heng}, K. 2018, \aj,
  155, 150

\bibitem[{{Menou} \& {Rauscher}(2009)}]{MenouRauscher2009}
{Menou}, K., \& {Rauscher}, E. 2009, \apj, 700, 887

\bibitem[{Mishchenko {et~al.}(1999)Mishchenko, Dlugach, Yanovitskij, \&
  Zakharova}]{Mishchenko1999}
Mishchenko, M.~I., Dlugach, J.~M., Yanovitskij, E.~G., \& Zakharova, N.~T.
  1999, Journal of Quantitative Spectroscopy and Radiative Transfer, 63, 409

\bibitem[{{Moreno} \& {Sedano}(1997)}]{Moreno1997heating}
{Moreno}, F., \& {Sedano}, J. 1997, \icarus, 130, 36

\bibitem[{{Moses} {et~al.}(2013){Moses}, {Madhusudhan}, {Visscher}, \&
  {Freedman}}]{Moses2013}
{Moses}, J.~I., {Madhusudhan}, N., {Visscher}, C., \& {Freedman}, R.~S. 2013,
  \apj, 763, 25

\bibitem[{{Munoz} \& {Isaak}(2015)}]{MnI2015}
{Munoz}, G., \& {Isaak}, K.~G. 2015, Proceedings of the National Academy of
  Science, 112, 13461

\bibitem[{{Oreshenko} {et~al.}(2016){Oreshenko}, {Heng}, \&
  {Demory}}]{Oreshenko2015}
{Oreshenko}, M., {Heng}, K., \& {Demory}, B.-O. 2016, \mnras, 457, 3420

\bibitem[{{Parmentier} {et~al.}(2016){Parmentier}, {Fortney}, {Showman},
  {Morley}, \& {Marley}}]{Parmentier2016}
{Parmentier}, V., {Fortney}, J.~J., {Showman}, A.~P., {Morley}, C., \&
  {Marley}, M.~S. 2016, \apj, 828, 22

\bibitem[{{Parmentier} {et~al.}(2013){Parmentier}, {Showman}, \&
  {Lian}}]{Parmentier2013}
{Parmentier}, V., {Showman}, A.~P., \& {Lian}, Y. 2013, \aap, 558, A91

\bibitem[{{Polichtchouk} {et~al.}(2014){Polichtchouk}, {Cho}, {Watkins},
  {Thrastarson}, {Umurhan}, \& {de la Torre Ju{\'a}rez}}]{Chogroup2014}
{Polichtchouk}, I., {Cho}, J.~Y.-K., {Watkins}, C., {et~al.} 2014, \icarus,
  229, 355

\bibitem[{Powell {et~al.}(2018)Powell, Zhang, Gao, \&
  Parmentier}]{powell2018formation}
Powell, D., Zhang, X., Gao, P., \& Parmentier, V. 2018, The Astrophysical
  Journal, 860, 18

\bibitem[{Rauscher \& Menou(2010)}]{RauscherMenou2010}
Rauscher, E., \& Menou, K. 2010, The Astrophysical Journal, 714, 1334

\bibitem[{{Rauscher} \& {Menou}(2012)}]{RauscherMenou2012}
{Rauscher}, E., \& {Menou}, K. 2012, \apj, 750, 96

\bibitem[{{Roman} \& {Rauscher}(2017)}]{Roman&Rauscher2017}
{Roman}, M., \& {Rauscher}, E. 2017, \apj, 850, 17

\bibitem[{{Rossow}(1978)}]{Rossow1978}
{Rossow}, W.~B. 1978, \icarus, 36, 1

\bibitem[{Showman {et~al.}(2009)Showman, Fortney, Lian, Marley, Freedman,
  Knutson, \& Charbonneau}]{showman2009atmospheric}
Showman, A.~P., Fortney, J.~J., Lian, Y., {et~al.} 2009, The Astrophysical
  Journal, 699, 564

\bibitem[{{Sing} {et~al.}(2016){Sing}, {Fortney}, {Nikolov}, {Wakeford},
  {Kataria}, {Evans}, {Aigrain}, {Ballester}, {Burrows}, {Deming},
  {D{\'e}sert}, {Gibson}, {Henry}, {Huitson}, {Knutson}, {Lecavelier Des
  Etangs}, {Pont}, {Showman}, {Vidal-Madjar}, {Williamson}, \&
  {Wilson}}]{SingNature2016}
{Sing}, D.~K., {Fortney}, J.~J., {Nikolov}, N., {et~al.} 2016, \nat, 529, 59

\bibitem[{{Stevenson} {et~al.}(2014){Stevenson}, {D{\'e}sert}, {Line}, {Bean},
  {Fortney}, {Showman}, {Kataria}, {Kreidberg}, {McCullough}, {Henry},
  {Charbonneau}, {Burrows}, {Seager}, {Madhusudhan}, {Williamson}, \&
  {Homeier}}]{Stevenson2014}
{Stevenson}, K.~B., {D{\'e}sert}, J.-M., {Line}, M.~R., {et~al.} 2014, Science,
  346, 838

\bibitem[{{Stevenson} {et~al.}(2017){Stevenson}, {Line}, {Bean}, {D{\'e}sert},
  {Fortney}, {Showman}, {Kataria}, {Kreidberg}, \& {Feng}}]{Stevenson2017}
{Stevenson}, K.~B., {Line}, M.~R., {Bean}, J.~L., {et~al.} 2017, \aj, 153, 68

\bibitem[{Toon {et~al.}(1989)Toon, McKay, Ackerman, \& Santhanam}]{Toon1989}
Toon, O.~B., McKay, C., Ackerman, T., \& Santhanam, K. 1989, Journal of
  Geophysical Research: Atmospheres, 94, 16287

\bibitem[{{Wakeford} {et~al.}(2017){Wakeford}, {Visscher}, {Lewis}, {Kataria},
  {Marley}, {Fortney}, \& {Mandell}}]{Wakeford2017}
{Wakeford}, H.~R., {Visscher}, C., {Lewis}, N.~K., {et~al.} 2017, \mnras, 464,
  4247

\bibitem[{{Webber} {et~al.}(2015){Webber}, {Lewis}, {Marley}, {Morley},
  {Fortney}, \& {Cahoy}}]{Webber2015}
{Webber}, M.~W., {Lewis}, N.~K., {Marley}, M., {et~al.} 2015, \apj, 804, 94

\bibitem[{{West} {et~al.}(1986){West}, {Strobel}, \& {Tomasko}}]{West1986}
{West}, R.~A., {Strobel}, D.~F., \& {Tomasko}, M.~G. 1986, \icarus, 65, 161

\end{thebibliography}
\bibliographystyle{apj}

\appendix
\label{appendix:a}
\section{Particle Scattering Parameters}
Our radiative transfer modeling of clouds required specifying parameters to describe the scattering and absorption of radiation, as listed in Table 2. These included the single scattering albedo (${\varpi_0}$), asymmetry parameter (${g_0}$), and the extinction efficiency (${Q_e}$) for each layer of particles. Optical depths per bar (${\tau_a^*}$) were computed based on the assumed molecular abundances (${\chi_{g}}$), the fraction of vapor condensed (${f}$), particles sizes, particle densities (${\rho}$), and computed extinction efficiencies.  Particles properties were evaluated at 650 nm and 5 $\mu$m, representing the two channels (visible and infrared) of our double-gray, two-stream radiative transfer scheme, for 0.2 $\mu$m particles using Mie theory and compositionally appropriate refractive indices.

\begin{deluxetable}{lccccc}[b]
\tabletypesize{\footnotesize}
\tablecolumns{5} 
\tablewidth{0pt}
 \tablecaption{ Aerosol Scattering Parameters}
 \label{table:scatparams}
 \tablehead{
 \colhead{Parameter} & \colhead{MgSiO${_3}$} & \colhead{Fe} & \colhead{Al${_2}$O${_3}$} & \colhead{MnS} }
 \startdata 
Mole fraction, ${\chi_{g}}$ & $3.26\times10^{-5}$ & $2.94\times10^{-5}$ & $2.77\times10^{-6}$ & $3.11\times10^{-7}$\\
Particle density, ${\rho}$ ( g/cm$^3$) & 3.2 & 7.9 & 4.0 & 4.0 &\\
Refractive index, visible& 1.7 + $i1\times10^{-4}$ & 2.5 + $i3.5$ & 1.7 + $i4\times10^{-2}$ & 2.9 + $i3\times10^{-4}$ \\
Refractive index, IR & 1.7 + $i6\times10^{-3}$ & 4.3+ $i10.5$ & 1.7 + $i2\times10^{-2}$ &2.9 + $i1\times10^{-9}$\\
\hline
{}&\multicolumn{4}{c} {Visible ${\vert}$ IR} \\
Single scattering albedo, ${\varpi_0}$ & 1.00 ${\vert}$ 0.16 & 0.69 ${\vert}$ 0.24 & 0.87 ${\vert}$ 0.03 & 1.00 ${\vert}$ 1.00 \\
Asymmetry parameter, ${g_0}$ & 0.65 ${\vert}$ 0.03 & 0.44 ${\vert}$ -0.17 & 0.66 ${\vert}$ 0.02 & 0.27 ${\vert}$ 0.06 \\
Extinction efficiency, ${Q_e}$ & 2.76 ${\vert}$ 0.01 & 2.98 ${\vert}$ 0.14 & 3.17 ${\vert}$ 0.01 & 2.93 ${\vert}$ 0.02 \\
{ }Thinner Cloud (${f}$=0.01): &&&& \\
Optical depth per bar, ${\tau_a^*}$ & 1143.8 ${\vert}$ 3.0 & 219.6 ${\vert}$ 11.9 & 38.9 ${\vert}$ .23 & 6.9 ${\vert}$ 0.05\\
{  }Thicker Cloud (${f}$=0.10): &&&&\\
Optical depth per bar, ${\tau_a^*}$ & 11438.0 ${\vert}$ 29.7 & 2196.0 ${\vert}$ 118.6 & 389.0 ${\vert}$ 2.3 & 69.0 ${\vert}$ 0.5\\
\enddata
 \tablecomments{Parameters were computed using a Mie theory code of M. I. Mishchenko \citep{deRooij1984,Mishchenko1999} assuming a log normal size distribution with the effective mean particle radii of 0.2 $\mu$m and a variance of 0.1 $\mu$m. Complex indices of refraction were taken from \cite{KitzmannHeng2018}. Wavelength dependent parameters were evaluated at 650 nm (vis) and 5.0 $\mu$m (IR).}
 \label{table:scatparamssmall}
\end{deluxetable}

\end{document}